\documentclass[fleqn,usenatbib]{mnras}
\usepackage{newtxtext,newtxmath}

\usepackage[T1]{fontenc}
\usepackage{amsmath}
\usepackage{graphicx}
\usepackage{lastpage}
\usepackage[dvipsnames]{xcolor}   
\usepackage[super]{nth}
\usepackage{microtype}
\usepackage{booktabs, tabularx}
\usepackage{supertabular}
\usepackage{physics}
\usepackage{enumitem}
\usepackage[math]{cellspace}

\usepackage{hyperref}
\definecolor{royalblue}{rgb}{0.25, 0.41, 0.88}
\definecolor{darkviolet}{rgb}{0.58, 0.0, 0.83}
\newcommand{\nq}[1]{%
	\begin{tabular}{@{}c@{}}\strut#1\strut\end{tabular}%
}

\title[Relic radiation from primordial tensor modes]{Towards a reliable calculation of relic radiation from primordial gravitational waves}

\author[W. Giar\`e, M. Forconi, E. Di Valentino, A. Melchiorri]{
William Giar\`e,$^{1,2,4}$\thanks{E-mail: w.giare@sheffield.ac.uk }
Matteo Forconi,$^{2,3}$\thanks{E-mail: matteo.forconi@roma1.infn.it }
Eleonora Di Valentino,$^{4}$\thanks{E-mail: e.divalentino@sheffield.ac.uk }
Alessandro Melchiorri,$^{2,3}$\thanks{E-mail: alessandro.melchiorri@roma1.infn.it }
\\
$^{1}$ Galileo Galileo Institute for theoretical physics, Centro Nazionale INFN di studi avanzati, Largo Enrico Fermi 2,  I-50125, Firenze, Italy\\
$^{2}$ Istituto Nazionale di Fisica Nucleare (INFN), Sezione di Roma, P.le A. Moro 2, I-00185, Roma, Italy\\
$^{3}$ Physics Department Universit\`a di Roma ``La Sapienza'', Ple Aldo Moro 2, 00185, Rome, Italy\\
$^{4}$ School of Mathematics and Statistics, University of Sheffield, Hounsfield Road, Sheffield S3 7RH, United Kingdom}

\pubyear{2022}

\begin{document}
\label{firstpage}
\pagerange{\pageref{firstpage}--\pageref{lastpage}}
\maketitle

\begin{abstract} 
Inflationary gravitational waves, behaving as additional radiation in the Early Universe, can increase the effective number of relativistic species ($N_{\rm eff}$) by a further correction that depends on the integrated energy-density in gravitational waves over all scales. This effect is typically used to constrain (blue-tilted) models of inflation in light of the bounds resulting from the Big Bang Nucleosynthesis. In this paper, we recompute this contribution, discussing some caveats of the state-of-the-art analyses. Through a parametric investigation, we first demonstrate that the calculation is dominated by the ultraviolet frequencies of the integral and therefore by the behavior of the tensor spectrum on scales corresponding to modes that cross the horizon very close to the end of inflation, when the slow-roll dynamics breaks down and the production of gravitational waves becomes strongly model dependent. Motivated by these results, we realize a theoretical Monte Carlo and, working within the framework of the Effective Field Theory of inflation, we investigate the observable predictions of a very broad class of models. For each model, we solve a system of coupled differential equations whose solution completely specifies the evolution of the spectrum up to the end of inflation. We prove the calculation of $\Delta N_{\rm eff}^{\rm GW}$ to be remarkably model-dependent and therefore conclude that accurate analyses are needed to infer reliable information on the inflationary Universe. 
\end{abstract}


\section{Introduction} 
\label{sec.1}

According to our current theory of the Early Universe, a phase of almost de-Sitter expansion known as cosmological Inflation~\citep{Guth:1980zm} is expected to drive the Universe towards homogeneity and flatness, setting the appropriate initial conditions for the subsequent Hot Big Bang Theory evolution and providing a compelling mechanism to explain the physical origin of the observed anisotropies in the Cosmic Microwave Background (CMB) radiation. 

A \textit{unique} prediction of inflation theory is the existence of Primordial Gravitational Waves (PGWs), tensor perturbations on super-horizon scales sourced by a super-adiabatic amplification of zero-point quantum fluctuations during inflation~\citep{Starobinsky:1980te,Linde:1981mu,Vilenkin:1983xq,Ungarelli:2005qb,Guzzetti:2016mkm}. Their detection would provide direct evidence for inflation, opening an inestimable observational windows on fundamental physics. For this reason, significant experimental efforts have been devoted to the search for primordial tensor modes, above all by looking for B-modes polarization on large angular scales in the Cosmic Microwave Background angular power spectra~\citep{Baumann:2014cja,Kamionkowski:2015yta}. Nevertheless, despite the best efforts, a detection of primordial tensor perturbations is still missing and only upper bounds can be inferred by current data~\citep{BICEP:2021xfz,Akrami:2018odb}. 

More precisely, within the simplest slow-roll scenario (where inflation is achieved by means of a single scalar field minimally coupled to gravity) the power spectrum of primordial tensor perturbations around the CMB scales can be well described by a two-parameter power-law parameterization:
\begin{equation}
\ln \mathcal P_{\rm T}(k)=\ln(r\,A_{\rm s}) + n_{\rm T}\,\ln(k/k_{\star}).
\label{PL}
\end{equation}
The first parameter, \textit{i.e.} the tensor amplitude $A_{\rm T}\doteq r\,A_{\rm s}$, is currently constrained to\footnote{We recall that $A_{\rm s}\simeq 2.1\times 10^{-9}$ is the amplitude of primordial scalar perturbations~\citep{Planck:2018jri}.} $r<0.032$ at 95\%CL~\citep{Tristram:2021tvh} when \textit{Planck}~\citep{Planck:2020olo} and BK18~\citep{BICEP:2021xfz} datasets are combined, together with BAO~\citep{eBOSS:2020yzd} and CMB lensing~\citep{Planck:2018lbu}. Hopefully, in the upcoming decade, new CMB experiments such as BICEP3~\citep{BICEP3}, CLASS~\citep{CLASS} , SPT-3G~\citep{SPT-3G}, Advanced ACTPol~\citep{ACTPol}, LiteBIRD~\citep{LBIRD} and CMB-S4~\citep{CMB-S4} should reach a better sensitivity $r \sim 0.001$, possibly leading to the first detection of B-mode polarization.
As concerns the second parameter, \textit{i.e.}, tensor tilt $n_{\rm T}\doteq d\ln \mathcal P_{\rm T}/d\ln k$, within the simplest single-field slow-roll framework, its value is fully determined by the slow-roll consistency relation $n_{\rm T}=-r/8$ that implies an almost scale-invariant slightly red-tilted spectrum. However this relation can be violated in many non-standard realizations of inflation such as in modified gravity theories~\citep{Baumann:2015xxa,Odintsov:2020ilr,Giare:2020plo,Oikonomou:2021kql,Odintsov:2022cbm}, in multi-fields inflationary models~\citep{Namba:2015gja,Peloso:2016gqs,Pi:2019ihn,Ozsoy:2020ccy}, or from trans-Planckian Physics~\citep{Ashoorioon:2014nta,Ashoorioon:2005ep}. Depending on the underlying phenomenology, the tensor tilt can range from being red ($n_{\rm T}<0$) to blue ($n_{\rm T}>0$), see e.g.~\citep{Stewart:2007fu, Mukohyama:2014gba,Giovannini:2015kfa,Giovannini:2018dob,Giovannini:2018nkt,Giovannini:2018zbf,Giare:2020vhn,Baumgart:2021ptt} and the references therein. As a result, constraining the tensor tilt (and in general the shape of the tensor spectrum) without any underlying assumption is crucial for testing new physics and the standard slow-roll scenario~\citep{Franciolini:2018ebs,DEramo:2019tit,Giare:2019snj,Caldwell:2018giq,Clarke:2020bil}.

Relaxing the slow-roll consistency relation, the analysis of the CMB data only weakly constrains the tensor tilt to $-0.55<n_{\rm T}<2.54$ at 95\% CL~\citep{Akrami:2018odb}. However, important improvements in the upper limit can be achieved by exploiting other CMB-independent observables. 
For instance, along with B-modes polarization, primordial tensor fluctuations may contribute also to the stochastic background of gravitational waves (SGWB), the analogous of CMB for gravitational waves~\citep{Caprini_2018}. Interestingly, if the spectrum is enough blue-tilted, according to Eq.\eqref{PL} the inflationary contribution should be much amplified on scales of direct gravitational wave detection so that we can use data from ground-based interferometers such as LIGO and VIRGO to infer constraints on $n_{\rm T}$. These experiments set an upper bound on the fraction of the energy-density of the Universe in gravitational radiation $\Omega_{\rm GW}\lesssim 10^{-7}$ ~\citep{LIGO_SGWB-2017,LIGO_SGWB-2019} in the frequency range $f\in\left(20\,\rm{-}\,85.8\right)$ Hz (which corresponds to the wave-number range  $k_{\rm LV} \in \left(1.3\,\rm{-}\,5.5\right)\times 10^{16} \,\rm{Mpc}^{-1}$), leading to a more stringent upper limit  $n_T < 0.52$ at 95\% CL ~\citep{Akrami:2018odb}. While this approach is largely used in the literature, it should be noted that these bounds are obtained by extrapolating the relation \eqref{PL} on frequencies (those probed by GWs experiments) where it is not granted that the spectrum still follows a power-law behavior. Indeed, high wave-numbers $k$ correspond to modes that exit the horizon relatively close to the end of inflation where the spectrum may strongly depend on the higher-order terms in Eq.~\eqref{PL}~\citep{Giare:2020vhn} and therefore on the specific form of the inflationary potential~\citep{Kinney:2021nje}, making it extremely difficult to derive reliable model-independent bounds on the tensor-tilt.

Another interesting possibility to gain constraining power on blue-tilted models of inflation is to study the effects induced by PGWs in the early Universe, before the recombination epoch. Behaving as extra radiation, a sizable amount of tensor perturbations may significantly contribute to the energy budget of the Early Universe, increasing the effective number of relativistic species $N_{\rm eff}$ by a further contribution~\citep{Maggiore:1999vm}
\begin{equation}
\Delta N_{\rm eff}^{\rm GW} \simeq \frac{h_0^2}{5.6\times10^{-6}}\left(\frac{1}{24\,z_{\rm eq}}\right) \int_{f_{\rm min}}^{f_{\rm max}} \frac{\mathrm{d}f}{f}\, \mathcal P_{\rm T}(f)
\label{Int1}
\end{equation}
that depends on the integrated energy-density in gravitational waves over all scales and that exponentially grows when $n_{\rm T}>0$, see also \hyperref[Appendix-A]{Appendix A} for a detailed derivation. So, in principle, we can use the Big Bang Nucleosynthesis (BBN) limit on additional radiation ($\Delta N_{\rm eff}\lesssim 0.4$) to infer constraints on blue-tilted models of inflation. Also this approach is largely followed in literature, leading to a limit $n_{\rm T}\lesssim 0.4$ that is more or less of the same order as those inferred by gravitational wave experiments, see \textit{e.g.} Refs.\citep{Allen:1997ad,Smith:2006nka,Boyle:2007zx,Kuroyanagi:2014nba,Ben-Dayan:2019gll,Aich:2019obd,Cabass:2015jwe}.

In the present work, we would like to focus a bit closer on this latter scenario. In 
\autoref{sec.3} we review the state-of-the-art analyses, outlining some important caveats and showing that the results share the same caveats discussed so far. Also in this case the largest inflationary contributions to the effective number of relativistic species come from tensor modes that exit the horizon very close to the end of inflation, precisely when the slow-roll approximation is no longer valid and the power-law parametrization breaks down. Consequently, any calculation becomes model-dependent and accurate analyses are needed to correctly estimate the relic radiation resulting from primordial tensor modes. To prove this point further and confer additional physical meaning to our findings, in \autoref{sec.4} we explicitly compute the energy budget of the Universe in several general Effective Field Theory (EFT) realizations of (blue and red) inflation. By integrating a set of differential equations we correctly predict the evolution of the spectrum (and all the other dynamical quantities) over the different cosmic epochs and scales. Finally, we present our conclusion in \autoref{sec.5}.

\section{Parametric Analysis}
\label{sec.3}

\subsection{State-of-the-art analyses}
\label{sec.3.1}
According to Eq.~\eqref{Int1} the contribution of inflationary tensor anisotropies to the effective number of relativistic degrees of freedom in the early Universe will depend on (i) the frequency range $f\in[f_{\rm min}\,,\,f_{\rm max}]$ over which the integral runs and (ii) the (parametrization of) primordial tensor spectrum. 

(i) The choice of the frequency range on which the integral runs is quite debated. In particular, the infrared cutoff can be safely set to $f_{\rm min} = 10^{-10}\,\rm{Hz}$ which approximately corresponds to the size of the comoving horizon at the time of BBN~\citep{Cabass:2015jwe,Pritchard:2004qp,Smith:2006nka}. Conversely, the ultraviolet cutoff is more arbitrary. Being primordial gravitational waves produced during inflation, we expect an ultraviolet cutoff of the size of the horizon at the end of inflation~\citep{Meerburg:2015zua} (as PGWs with smaller wavelengths cannot be produced). Anyway, the size of the horizon at the end of inflation depends on the reheating temperature $T_{\rm RH}$ at the end of inflation. Assuming an almost GUT-scale inflation and an instant reheating we can set $T_{\rm RH}\sim 10^{15}\,\rm{GeV}$ which corresponds to $k_{\rm end}\sim 10^{23}\,\rm{Mpc}^{-1}$ and thus $f_{\rm max}\simeq 10^{8}\,\rm{Hz}$~\citep{Cabass:2015jwe}. Nevertheless, inflationary models with (very) lower reheating temperatures $T_{\rm RH}\sim 10^{10} - 100\, \rm{GeV}$ have been proposed in the literature (see e.g., Refs.~\citep{Kawasaki:1999na,Kawasaki:2000en,Giudice:2000dp,Giudice:2000ex,Hannestad:2004px,Khoury:2011ii,Hasegawa:2019jsa,Hasegawa:2020ctq,Carenza:2021ebx,Freese:2017ace,Litsa:2020rsm}) and, although such scenarios are typically not easy to realize, in these models the ultraviolet cutoff may be much smaller, limiting the high-frequency contributions in the integral \eqref{Int1}, see also Refs.~\citep{Vagnozzi:2020gtf,Benetti:2021uea}. 

(ii) The main purpose of this section is to study the dependence of the integral \eqref{Int1} from the parametrization used for the primordial tensor spectrum $\mathcal P_{ \rm T}$. The common practice in literature is to assume a power-law tensor spectrum given by Eq.~\eqref{PL} over the whole range of integration so that the integral \eqref{Int1} can be easily solved analytically:
\begin{align}
\Delta N_{\rm eff}^{\rm GW}& \simeq \frac{h_0^2}{5.6\times10^{-6}}\left(\frac{ r A_s}{24\,z_{\rm eq}}\right) \frac{1}{n_{\rm T}} \left[\left(\frac{f}{f_{\star}}\right)^{n_{\rm T}} \right]^{f_{\rm max}}_{f_{\rm min}}
\label{eq:limit1}
\end{align}
Interestingly, a blue tensor tilt exponentially amplifies the GWs production on ultraviolet frequencies - that therefore we expect to contribute mostly in  Eq.~\eqref{Int1} - possibly leading to a sizable $\Delta N_{\rm eff}$ from PGWs. As we already discussed in the introduction, this effect is commonly used in literature to bound blue-tilted models of inflation, with several Implications also for gravitational waves observations~\citep{Vagnozzi:2020gtf,Benetti:2021uea,Vagnozzi:2022qmc} and fundamental physics~\citep{Calcagni:2020tvw}. For instance, assuming a GUT scale inflation ($f_{\rm max}\sim 10^{8}$ Hz) and a tensor amplitude $r\sim 0.001$, it is easy to see that the BBN limit on the the effective number of relativistic species ($\Delta N_{\rm eff}\lesssim 0.4$) is naturally translated into a limit $n_{\rm T}\lesssim 0.4$ by Eq.~\eqref{eq:limit1}, see also \autoref{fig:figure1} and \hyperref[sec.Appendix-B]{Appendix B} where an updated analysis of the observational constraints resulting from the BBN is carried out. We devote the rest of this section to studying how much robust these bounds are.  

\subsection{Next-to-leading order parameterization}
\label{sec.3.2}
A first naive consideration is that the above mentioned result is derived assuming the tensor tilt to be exactly constant under the whole range of integration. Typically, in physical models of inflation where the tensor tilt can acquire such large positive values, it may also acquire a non-negligible scale dependence ~\citep{Giare:2020vhn,Giare:2020plo}. Therefore, a first attempt to question the strength of this result is to study what happens extending the power low relation \eqref{PL} to its next-to-leading order generalization
\begin{equation}
\ln \mathcal P_{\rm T}(k)=\ln(r\,A_{\rm s}) + n_{\rm T}\,\ln(k/k_{\star}) + \alpha_{\rm T}\,\ln^2(k/k_{\star})
\label{PL2}
\end{equation}
where we parametrize the scale dependence of the tensor tilt by including its running $\alpha_{\rm T}\doteq dn_{\rm T} / d\ln k$.

\begin{figure}[t]
 	 \centering
 	 \includegraphics[width=\columnwidth]{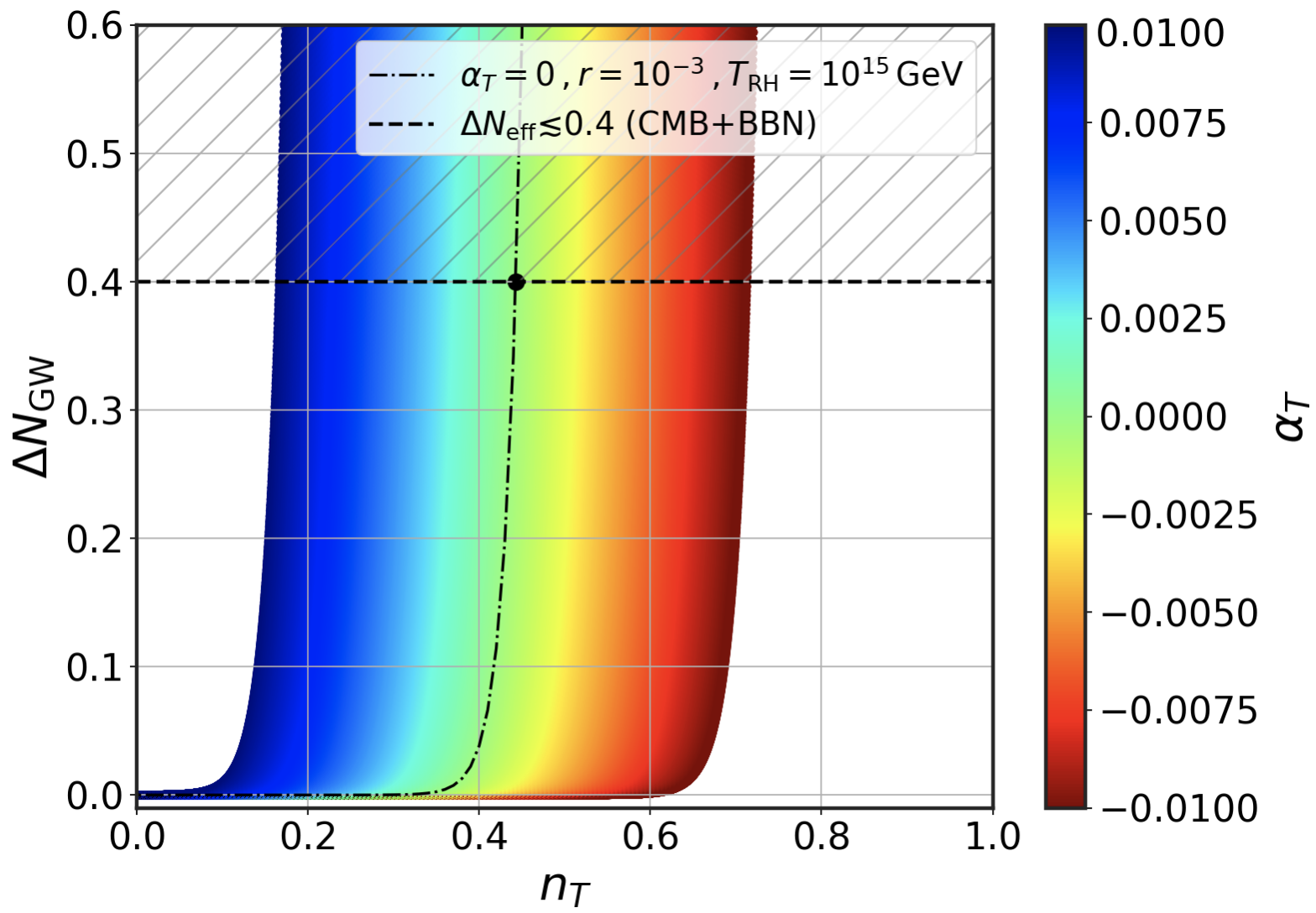}
 	 \caption{\small Inflationary tensor mode contribution to the effective number of relativistic degrees of freedom as a function of the tensor tilt and its running $\alpha_{\rm T}$. The black dashed line represents the contribution for $\alpha_{\rm T}=0$ while the horizontal dashed line represents the limit on additional radiation from the BBN bounds.}
 	 \label{fig:figure1}
 \end{figure}

In \autoref{fig:figure1}, we show the effect of a relatively small running of the tensor tilt on the calculation of $\Delta N_{\rm eff}^{\rm GW}$ finding that it can significantly change the results and so lead to a much tighter (relaxed) constraint on $n_{\rm T}$ represented by the horizontal dashed line in the figure. We postpone a rigorous analysis of the effects of a running of the tensor tilt on the observational constraints resulting from the BBN to \hyperref[sec.Appendix-B]{Appendix B}. Here we point out that a positive (negative) $\alpha_{\rm T}$ amplifies (suppresses) the power spectrum on high frequency and its contributions in the integral \eqref{Int1}, providing another important clue that properly accounting for the ultraviolet behavior of the tensor spectrum may be crucial in the calculation of $\Delta N_{\rm eff}^{\rm GW}$. In this regard, we notice that modes with frequency $f=k/2\pi$ will cross horizon $N_{k}$ e-folds before the end of inflation, where $N_{k}$ is given by~\citep{Martin:2013tda,Kinney:2021nje}
\begin{align} 
\nonumber N_{k}\simeq&-\ln \left(\frac{k}{a_{0} H_{0}}\right)+\ln \left(\frac{H_{\star}}{H_{\rm {end}}}\right)-\frac{2}{3}\ln\left(\frac{T_{\rm RH}}{\Lambda}\right) \\ &\nonumber +\ln \left(\frac{T_{\rm{RH}}}{T_{\rm {\rm{eq}}}}\right)+\frac{1}{3} \ln \left(\frac{g_{* S}\left(T_{R H}\right)}{g_{* S}\left(T_{\rm{eq}}\right)}\right) \\ &+\ln \left(\frac{a_{\rm{eq}}H_{\rm{eq}}}{a_0 H_0}\right) .
\label{Nk}
\end{align}
In the equation above $a_0H_0= 2.248 \times 10^{-4} \rm{Mpc}^{-1}$ is the inverse of the comoving horizon size in the current Universe, $H_\star$ is the value of the Hubble parameter at the horizon exit, $H_{\rm end}$ is the Hubble parameter at the end of inflation, $\Lambda$ is the energy scale of inflation and the subscript "eq" denotes quantities evaluated at matter-radiation equality. Assuming a standard $\Lambda$CDM cosmology, we have $\ln \left[(a_{\rm{eq}} H_{\rm{eq}}) /(a_0 H_0) \right]\simeq 3.8$ and $T_{\rm eq}\simeq 8\times10^{-10}$ GeV~\citep{Martin:2013tda,Planck:2018jri,Forconi:2021que}. Approximating $H_{\rm end}\simeq H_{\star}$ and recalling that the energy scale of inflation can be related to the amplitude of tensor perturbations as $\Lambda\simeq r^{1/4}\times 3.3\times 10^{16}\rm{GeV}$, we can simplify Eq.~\eqref{Nk} to 
\begin{align} 
N_{k}\simeq 61-\ln \left(\frac{k}{a_{0} H_{0}}\right)+\frac{1}{3}\ln\left(\frac{T_{\rm RH}}{10^{15}\,\rm{GeV}}\right) +\frac{1}{6}\ln\left(r\right)  \,.
\end{align}
Therefore the "high frequencies" in the integral \eqref{Int1} we are referring to, correspond to tensor modes that exit the horizon extremely close to the end of inflation ($N_k\lesssim 2$ for $k\gtrsim 10^{21}\,\rm{Mpc^{-1}}$ and $T_{\rm RH}\sim 10^{15}$ GeV and $r\sim 10^{-3}$). This is precisely where, at least in the simplest inflationary scenarios, the potential decreases very rapidly to approach its minimum, and the slow-roll dynamics breaks down. As pointed out in Refs.~\citep{Kinney:2021nje,Giare:2020vhn}, it is not sure at all that a power-law parameterization (or even its next-to-leading order generalization) holds - even approximately - on such frequencies because the shape of the tensor spectrum will be strongly related to the shape of the inflationary potential. As a result, we argue the calculation of $\Delta N_{\rm eff}^{\rm GW}$ to be largely sensitive to the underlying model. 
 
\subsection{Higher-order stochastic reconstruction}
\label{sec.3.3}

\begin{figure}
 	 \centering
 	 \includegraphics[width=\columnwidth]{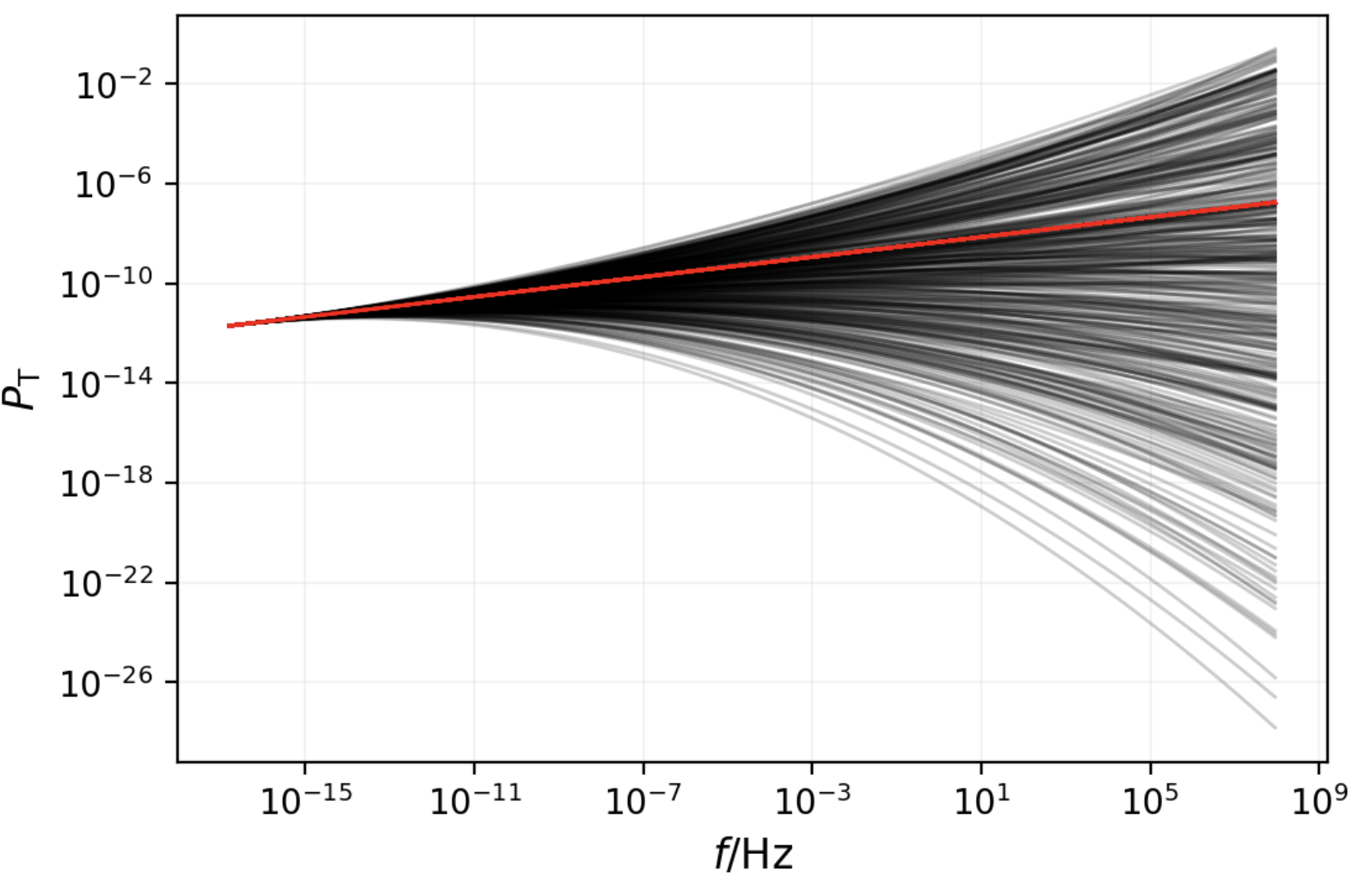}
 	 \caption{\small Examples of randomly generated tensor spectra (and the power-law extrapolation, red line) obtained by following the method outlined in \autoref{sec.3.3}.}
 	 \label{fig:figure2}
 \end{figure}

A more general approach to the problem can be obtained by expanding (the log of) the tensor spectrum as a series of powers 
\begin{align}
\ln \mathcal P_{\rm T}= \sum_{j=0}^{\infty} a_j (x-x_0)^j .
\label{EXP}
\end{align}
If we choose the CMB frequency as the center of the expansion ($x_0=\ln f_{\star}$) the coefficients $a_j$ can be trivially related to (the derivatives of) the tensor spectrum evaluated at the CMB scales. In particular, the tensor amplitude and the tensor tilt are simply given by
\begin{equation}
a_0= \ln(rA_S), \quad a_1=\frac{d\ln \mathcal P_{\rm T}}{ d\ln f}\equiv n_{\rm T}
\label{eq:a0a1}
\end{equation}
while the higher order coefficients are related to the higher order derivatives of the spectrum (or the tensor tilt) as: 
\begin{equation}
\quad a_{j>1}=\frac{1}{j!} \frac{d^j\ln \mathcal P_{\rm T}}{ d\ln^j f} = \frac{1}{j!} \frac{d^{j-1} n_{\rm T}}{d\ln^{j-1} f}.
\label{eq:aj}
\end{equation}
Notice that if we stop the sum expansion at $j=1$ or $j=2$, we exactly recover Eq.\eqref{PL} or Eq.\eqref{PL2}, respectively. Therefore including more and more terms in the sum will clearly guarantee a more accurate reconstruction of the tensor spectrum at $x \gg x_0$ since it employs also the other higher-order terms in the expansion. However, if we want to adopt this parameterization in the integral \eqref{Int1}, we need to make sure that this sum will actually converge on the frequencies over which the integration runs. Although this depends on the specific model of inflation, in most models the tensor spectrum is a slow-evolving regular function of the frequency so that it is reasonable to expect a global convergence. For instance, the simplest slow-roll scenario is characterized by a hierarchy of parameters $n_{\rm T}=\mathcal O(\epsilon)$ and $d^{j} n_{\rm T} /d\ln^{j} f\lesssim \mathcal O(\epsilon^{j+1})$. Assuming such a scaling, the sum convergence can be easily proved by evaluating the radius of convergence
\begin{equation}
\frac{1}{R}\doteq\lim_{j\to \infty} \left|\frac{a_{j+1}}{a_j}\right| = \lim_{j\to \infty} \left|\frac{\mathcal O(\epsilon)}{j+1}\right|= 0.
\end{equation}
So, in principle, we can adopt this parameterization to predict the value of the tensor spectrum at $x\gg x_0$. Anyway, in practice, all the arbitrariness of the method is encapsulated into the coefficients $\{a_j\}$. Ultimately, fixing their values is equivalent to fixing a specific model of inflation. Here we sample different inflationary models by randomly varying the coefficients $\{a_j\}$ as follows:
\begin{itemize}[leftmargin = *]

\item We fix the tensor amplitude\footnote{ Notice that $\Delta N_{\rm eff}^{\rm GW}$ can be easily obtain for any generic $r$ simply re-scaling the value obtained for $r=10^{-3}$ as
$$ \Delta N_{\rm eff}^{\rm GW}(r) = \left(\frac{r}{10^{-3}}\right) \bigg[\Delta N_{\rm eff}^{\rm GW}\bigg]_{r=10^{-3}} $$} on the CMB scales to $r\sim 10^{-3}$ (which is the target of the next CMB experiments) so that $a_0$ is always fixed by Eq.\eqref{eq:a0a1};

\item We let the tensor tilt randomly vary in the range $n_{\rm T}\in [-0.5,1]$ thus evaluating $a_1$ according to Eq.\eqref{eq:a0a1}. In this way, we can explore both blue and red tilted models\footnote{Notice that we are relaxing the slow-roll consistency relation between the tensor tilt and the tensor amplitude ($n_{\rm T}\ne -r/8$) and considering the two parameters as independent.};

\item We randomly choose the higher-order coefficients $\{a_{j>1}\}$ to be extremely small such that $a_1\gg a_{j}\gg a_{j+1}$. This is done by assuming the $j$-order derivative of the tensor spectrum in Eq.\eqref{eq:aj} to be a Gaussian distributed with mean $\mu=0$ and standard deviation $\sigma\simeq 10^{-2j}$. While this is clearly an arbitrary assumption, in this way we can be sure that the spectrum follows a power-law \eqref{PL} on the CMB scales ($x\simeq x_0$) where such terms remain in fact negligible. In addition, this ensures a fast convergence of the sum expansion on high frequencies ($x \gg x_0$) while granting a certain freedom.
\end{itemize}
Following this scheme we simulate $10^{6}$ different shapes of the tensor spectrum as functions of frequency up to the order $j=10$ in the sum expansion\footnote{We checked that maintaining this scaling for parameters, the 10th order is enough to capture any relevant correction to the tensor spectrum.}. Examples of the spectra obtained within this method are provided in \autoref{fig:figure2}, together with a simple leading order power-law approximation (red line).

Notice that we consider both negative and positive coefficients $\{a_j\}$, so that, on high frequencies, the spectrum can be either suppressed or amplified. Indeed, while in the simplest cases we expect suppression of power because of the rapid decrease of inflationary potential (see also the subsequent discussion in \autoref{sec.4}), in more elaborated scenarios it is in principle possible to build inflationary models with ultraviolet amplification of tensor perturbations~\citep{Oikonomou:2022ijs,Barrow:1993ad,Peng:2021zon,Ota:2022hvh,Odintsov:2022sdk,Baumgart:2021ptt}. As explained in the introduction, in this latter case we may end up with large amounts of GW on the small scales as those probed by Gravitational interferometers. Therefore, for all the simulated spectra, we also checked that the amplitude $\mathcal P_{\rm T}(k)$ remains consistent with the LIGO/VIRGO limit, keeping only the models able to satisfy observations. This is the reason why in \autoref{fig:figure2} we get much more suppressed spectra than amplified ones. From the same figure, we can also appreciate how the usual power-law parametrization is a precise approximation only at frequencies corresponding to the CMB scales (as required by construction) while important deviations are observed at higher frequencies, in spite of our efforts for keeping small the parameters $\{a_j\}$. 
  \begin{figure*}
 	 \centering
 	 \includegraphics[width=0.8\linewidth]{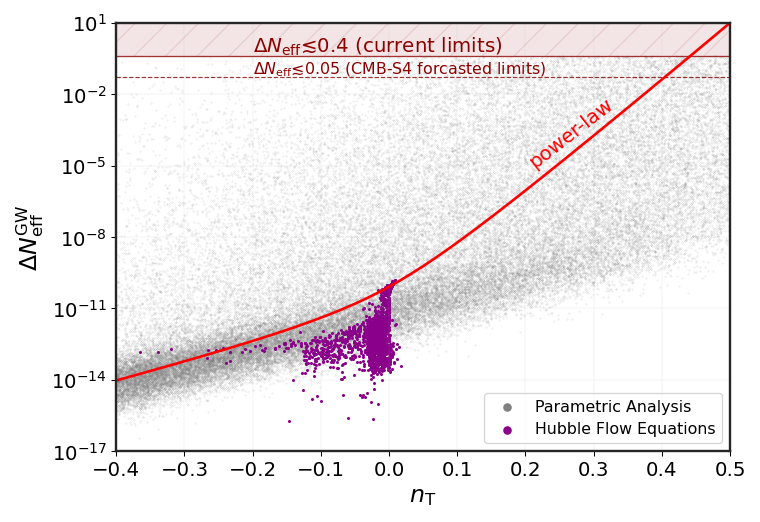}
 	 \caption{\small Primordial Gravitational Wave contribution to radiation energy-density in the early Universe parametrized as a correction to the effective number of relativistic species ($\Delta N_{\rm eff}^{\rm GW}$). All the inflationary models in the figure share the same tensor amplitude ($r\simeq 0.001$) and the same reheating temperature ($T_{\rm RH}\sim 10^{15} \text{GeV}$) but have different values of tensor tilt ($n_{\rm T}$) shown in the $x$-axis. The red thick line represents the predictions for $\Delta N_{\rm eff}^{\rm GW}$ inferred by extrapolating a power-law parameterization for the tensor spectrum (\autoref{PL}) over all frequencies. The gray dots represent the results of the parametric analysis carried out in \autoref{sec.3.3} where the spectrum is expanded as a sum of powers up to the 10th order (\autoref{EXP}) and randomly reconstructed. Finally, the magenta points represent the observable predictions of an ensemble of physical models randomly realized within the framework of the Effective Field Theory of inflation by means of a theoretical Monte Carlo. In this latter case, the spectrum is calculated by integrating a system of coupled differential equations (known as "Hubble Flow Equations"), as discussed in \autoref{sec.4}. The horizontal red band (dashed line) represents the current (future forecasted) observational limit on radiation.}
    \label{fig:figure3}
 \end{figure*}
 
Fixing the ultraviolet cutoff to $f_{\rm max}\simeq 10^{8}\,\rm{Hz}$, we numerically solve the integral \eqref{Int1} for all the different shapes of $\mathcal P_{\rm T}(f)$, thus computing the corresponding value of $\Delta N_{\rm eff}^{\rm GW}$. We ensure the computational relative error due to the numerical integration method to remain smaller than 1\%. In \autoref{fig:figure3} we show the results of our random analysis. Once again the red solid line represents the contribution $\Delta N_{\rm eff}^{\rm GW}$ obtained within the power-law parametrization \eqref{PL}. Instead, the gray dots represent the values of $\Delta N_{\rm eff}^{\rm GW}$ obtained by the numerical integration method of the randomly obtained tensor spectra. 

Despite the intrinsic aleatory nature of this method, we can certainly draw some general conclusions. First of all, as evident from \autoref{fig:figure2}, the high-frequency behavior of the tensor spectrum may become basically uncorrelated with the value of the tensor tilt on the CMB scales. This goes in the direction of previous analyses already discussed in the literature, see \textit{e.g.}, Refs.~\citep{Giare:2020vhn, Kinney:2021nje}. In addition, the results displayed in \autoref{fig:figure3} lead weight to our previous considerations according to which the value of $\Delta N_{\rm eff}^{\rm GW}$ may be strongly sensitive to the high-frequency contributions in the integral \eqref{Int1}. Since on such frequencies the spectrum becomes uncorrelated with the behavior of the tensor tilt on the CMB scale, these findings lead us to believe that the BBN limit on additional radiation can hardly constrain the tensor tilt itself, unless without a full understanding of the underlying model. It is worth noting that the equation Eq.~\eqref{EXP} amplifies higher order terms at ultraviolet frequencies. As a result, increasing the value of the ultraviolet cutoff ($f_{\rm max}$) will lead to larger contributions from non-linear terms in the integral~\eqref{Int1}. These larger contributions will cause greater dispersion in the gray points in \autoref{fig:figure3} and enhance differences in $\Delta N_{\rm eff}$, as noted in \citep{Vagnozzi:2020gtf}. This indicates that the parametric analysis is dependent, to some degree, on the choice of ultraviolet cutoff (which is fixed to $f_{\rm max}=10^{8}$ Hz in this case). Inflationary models that produce satisfying amounts of gravitational waves typically predict high-scale inflation, so for the values of tensor amplitude of interest to future experiments, a significant reduction in the ultraviolet cutoff is possible only within models with extremely low $T_{\rm RH}$. While these models are theoretically possible, they are very difficult to realize. In conclusion, this parametric analysis can be useful for pointing out potential limitations and weaknesses in current analyses, but a more reliable investigation of physical models of inflation and their respective contribution to the energy budget of the early Universe is needed. This will be the focus of the next section.



\section{Physical Analysis}
\label{sec.4}

The lesson we have learned from the parametric analyses detailed in the previous section is that the calculation of relic radiation from primordial gravitational waves depends crucially on the behavior of the primordial tensor spectrum at ultraviolet frequencies. Given that assuming a power-law continuously on all frequencies is not reliable~\citep{Giare:2020vhn, Kinney:2021nje}, the calculation of $\Delta N_{\rm eff}^{\rm GW}$ becomes unreliable in turn. Motivated by these results, in this section we want to provide a definitive evidence that this issue persists in solid theoretical framework of inflation, conferring physical meaning to our findings. In addition, we want to quantify the typical error resulting from extrapolating a power-law parameterization by going through a precise evaluation of the radiation energy-density for a reasonable range of different models and possibilities.

In order to investigate the observable predictions of a very broad class of inflationary models in the most general framework, we follow a methodology based on the so-called Hubble Flow Equation~\citep{Hoffman:2000ue,Kinney:2002qn,Easther:2002rw,Friedman:2006zt}. The Hubble Flow Equations were first introduced by Hoffman and Turner~\citep{Hoffman:2000ue} for the simplest single-field slow-roll case where it is straightforward to define an infinite hierarchy of slow-roll parameters that, starting from the Hubble parameter $H$ and its derivatives with respect to the field, completely specify the evolution of the main observable quantities during inflation. Since the integration of the equations yields a trajectory in slow-roll parameter space that can be ultimately interpreted as a model whose dynamics is a solution of the flow equations, solving numerically a truncated system of Hubble Flow Equations for a set of suitably defined initial conditions has been proposed as a sophisticated algorithm for generating large numbers of slow-roll inflationary models, without relying on the explicit form of the action~\citep{Kinney:2002qn}. 

Recently, in Ref.~\citep{Capurri:2020qgz} the method has been extended to the Effective Field Theory (EFT) framework of inflation to include a much broader class of beyond-standard inflationary models and explore a wide variety of possible high-energy corrections to the simplest slow-roll scenario. In this section, we follow this latter generalized approach to investigate in a more general and reliable way the actual contribution of inflationary tensor perturbations to the energy budget of the early Universe. We start reviewing the Hubble Flow Equations in the Effective Field Theory of Inflation, strictly following Ref.~\citep{Capurri:2020qgz}. Then we explain how we adapt this method to our investigation. Finally, we discuss the results.

\subsection{Hubble Flow Equations and EFT of Inflation}
The EFT of inflation~\citep{Cheung:2007st,Weinberg:2008hq} is a very general framework for describing fluctuations
around a quasi-de Sitter background. The general form of the effective action
in the comoving gauge reads
\begin{equation}
   \begin{aligned} S=& \int d^{4} x \sqrt{-g}\left[\frac{1}{2} M_{\mathrm{pl}}^{2} R-c(t) g^{00}-\Lambda(t)+\right.\\ &+\frac{1}{2 !} M_{2}(t)^{4}\left(g^{00}+1\right)^{2}+\frac{1}{3 !} M_{3}(t)^{4}\left(g^{00}+1\right)^{3}+\ldots \\ &\left.-\frac{\bar{M}_{1}(t)^{3}}{2}\left(g^{00}+1\right) \delta K_{\mu}^{\mu}-\frac{\bar{M}_{2}(t)^{3}}{2} \delta K_{\mu}^{\mu 2}\right.\\ &\left.-\frac{\bar{M}_{3}(t)^{3}}{2} \delta K_{\nu}^{\mu} \delta K_{\mu}^{\nu}+\ldots\right] \end{aligned}
   \label{eq:EFT_action}
\end{equation}
For the following discussion, it is useful to divide this action into two different blocks and analyze them separately. 

The fist important block is given by the first line of Eq.~\eqref{eq:EFT_action} that we rewrite below for convenience: 
\begin{equation}
S_{\rm{bg}}=\int d^{4} x \sqrt{-g}\left[\frac{1}{2} M_{\mathrm{pl}}^{2} R-c(t)g^{00}-\Lambda(t)\right]
\label{eq:background}
\end{equation}
It contains the standard Einstein-Hilbert action and terms that are linear perturbations around the background. Therefore, once that  the time-dependent coefficients $c(t)$ and $\Lambda (t)$ have been specified, this part of the action completely fixes the background evolution during inflation. Notice also that the evolution of the parameters $c(t)$ and $\Lambda (t)$ can be related to the evolution of the Hubble parameter by the Friedmann equations 
\begin{equation}
H^{2}=\frac{1}{3 M_{\mathrm{pl}}^{2}}[c(t)+\Lambda(t)] \quad \text { and } \quad \frac{\ddot{a}}{a}=-\frac{1}{3 M_{\mathrm{pl}}^{2}}[2 c(t)-\Lambda(t)]
\label{eq:lambda}
\end{equation}
so we need only two independent functions to fully characterize the background evolution that we choose to be $H(t)$ and $c(t)$, fixing $\Lambda(t)$ by Eq.~\eqref{eq:lambda}. Starting from Eq.~\eqref{eq:background}, we take a first step deriving the generalize Hubble Flow Equations for the background parameters. In analogy with the standard case, we take as our fundamental quantity the Hubble parameter as a function of inflaton field, $H(\phi)$. To switch from the time domain to field domain we can  exploit a relation betweeen the time-derivative of the field, $c(\phi)$ and $H(\phi)$ that follows from a combination of the Friedmann equation and the continuity equation, namely:
\begin{equation}
    \frac{\text{d}\phi}{\text{d}t}=-\frac{c(\phi)}{M_{\mathrm{pl}}^{2} H^{\prime}(\phi)}
\end{equation}
where, from now on, the prime indicates a derivative with respect to the field ($X^{\prime  } = d X / d\phi$). Using the relation above, it is easy to see that the slow roll parameter $\epsilon$ becomes:
\begin{equation}
    \epsilon=-\frac{\dot{H}}{H^{2}}=\frac{c(\phi)}{M_{\mathrm{pl}}^{2} H^{2}(\phi)}
\end{equation}
Starting from $\epsilon$, we can define the higher-order slow-roll parameters by iterated derivations:
\begin{equation}
    \begin{aligned} \eta(\phi)=& \frac{c(\phi)}{M_{\mathrm{pl}}^{2}} \frac{H^{\prime \prime}(\phi)}{H(\phi) H^{\prime 2}(\phi)} \\ & \vdots \\{ }^{l} \lambda(\phi) &=\left(\frac{c(\phi)}{M_{\mathrm{pl}}^{2}}\right)^{l}\left(\frac{1}{H(\phi)}\right)^{l}\left(\frac{1}{H^{\prime}(\phi)}\right)^{l+1} \frac{d^{l+1} H(\phi)}{d \phi^{l+1}} \end{aligned}
    \label{Htower}
\end{equation}
with $l \geq 2$ and $\eta(\phi)\equiv \, ^{1}\lambda(\phi)$. Notice however that, in contrast with the standard Hubble flow equations, now the evolution of $\epsilon$ and the other higher-order parameters will depend also on the additional unknown function $c(\phi)$. Therefore we need to define other new slow-roll parameters to describe the evolution of $c(\phi)$. Following the notation of Ref~\citep{Capurri:2020qgz} we introduce the parameter $\theta$
\begin{equation}
    \theta \equiv-\frac{\dot{c}}{H c}=\frac{1}{M_{\mathrm{pl}}^{2}} \frac{c^{\prime}(\phi)}{H(\phi) H^{\prime}(\phi)}
\end{equation}
and the the other higher-order parameters by taking iterated derivations 
\begin{equation}
   \begin{aligned} \kappa(\phi) &=\frac{1}{M_{\mathrm{pl}}^{2}} \frac{c^{\prime \prime}(\phi)}{H^{\prime 2}(\phi)} \\ & \vdots \\ ^{l}{\xi}(\phi) &=\left(\frac{c(\phi)}{M_{\mathrm{pl}}^{2}}\right)^{l}\left(\frac{1}{H(\phi)}\right)^{l-1}\left(\frac{1}{H^{\prime}(\phi)}\right)^{l+1} \frac{1}{c(\phi)} \frac{d^{l+1} c(\phi)}{d \phi^{l+1}} \end{aligned} 
   \label{Ctower}
\end{equation}
always with $l \geq 2$ and $\kappa(\phi)\equiv \, ^{1}\xi(\phi)$. An explicit calculation of the equations above lead to derive the generalized Hubble flow equation for the background parameters~\citep{Capurri:2020qgz}:
\begin{equation}
    \begin{cases}
    \frac{\text{d}\epsilon}{\text{d}N}=\epsilon\,\left( \theta - 2\epsilon \right)\\
    
    \frac{\text{d}\eta}{\text{d}N}= \eta\,\left( \theta - \epsilon -2\eta\right) +\,^{2}\lambda \\
    \vdots
    \\
    \frac{\text{d} \, ^{l}\lambda }{\text{d}N} = \, ^{l}\lambda \, \left[ l\left( \theta -\epsilon \right) -\left(l+1\right)\eta \right] + \, ^{l+1}\lambda
    \\
    \\
    \frac{\text{d}\theta}{\text{d}N}=\epsilon\,\kappa \, -\theta\,\left( \epsilon + \eta \right)\\
    \frac{\text{d} {\kappa} }{\text{d}N} = -2\kappa\eta + \, ^{2}\xi\\
    \vdots
    \\
    \frac{\text{d} \, ^{l}\xi }{\text{d}N} = \, ^{l}\xi \left[ \left(l-1\right)\left(\theta -\epsilon \right) -\left(l+1\right)\eta\right] + \, ^{l+1}\xi
    \end{cases}
    \label{eq:HFE_background}
\end{equation}
We stress that the integration of this system of coupled equations completely specifies the dynamics of the background during inflation.

The second block in the action~\eqref{eq:EFT_action}, involves the higher order operators that we have organized in powers of the number of perturbations and in terms of the
increasing number of derivatives:
\begin{equation}
\begin{aligned}
\Delta S = & \int d^{4} x \sqrt{-g}\left[ \sum_{n\geq 2} \frac{1}{n!}M_n(t)^{4}\left(g^{00}+1\right)^{n}\right. \\ &\left.-\frac{\bar{M}_{1}(t)^{3}}{2}\left(g^{00}+1\right) \delta K_{\mu}^{\mu}-\frac{\bar{M}_{2}(t)^{3}}{2} \delta K_{\mu}^{\mu 2}\right.\\ &\left.-\frac{\bar{M}_{3}(t)^{3}}{2} \delta K_{\nu}^{\mu} \delta K_{\mu}^{\nu}+\ldots\right]
\label{eq:DeltaS}
\end{aligned}
\end{equation}
These operators are turned on and off by the $M$ coefficients in the action, whose value will thus weight the relative effects. As we shall see, in their turn the coefficients $M$ can be related to physical quantities that can be in principle measured and constrained. Therefore, once we have reconstructed the background dynamics by solving the system~\eqref{eq:HFE_background}, it is useful to derive a further system
of equations to describe the evolution of the $M$ coefficients in Eq.~\eqref{eq:DeltaS} over that background. We can do so in a quite general and elegant way by noting that for any quantity described by a generic scalar function $Q(\phi)$, one can always define a slow-roll
parameter $\epsilon_Q$ as follows: 
\begin{equation}
    \epsilon_{Q}=-\frac{\dot{Q}}{H\,Q}=\frac{1}{M_{\mathrm{pl}}^{2}}\frac{c(\phi)}{H(\phi)\,H^{\prime}(\phi)}\,\frac{Q(\phi)}{Q^{\prime}(\phi)}
    \label{eq:epsQ}
\end{equation}
In analogy to the discussion for the background parameters, we define also the higher-order parameters for the quantity $Q(\phi)$ by taking its derivatives:

\begin{equation}
\begin{aligned} 
\rho_{Q}(\phi)&=\frac{1}{M_{\mathrm{pl}}^{2}} \frac{c(\phi)}{H^{\prime\,2}(\phi)}\frac{Q^{\prime \prime}(\phi)}{Q(\phi)}\\ & \vdots \\
{}^{l}\chi_{Q}(\phi)&=\left( \frac{c(\phi)}{M_{\mathrm{pl}}^{2}} \right)^l\,\left(\frac{1}{H(\phi)}\right)^{l-1}\,\left(\frac{1}{H^{\prime}(\phi)}\right)^{l+1}\,\frac{1}{Q} \frac{\text{d}^{l+1} \,Q}{\text{d}\phi^{l+1}}
\end{aligned} 
\end{equation}
again with $l\geq 2$ and $\rho_{Q}(\phi)\equiv{}^{1}\chi_{Q}(\phi)$. By explicitly computing these relations, we eventually get the system of Hubble flow equations for $Q(\phi)$:
\begin{equation}
\begin{cases}
\frac{\text{d} \epsilon_{Q}}{\text{d} N}= \epsilon_{Q}\left(\theta - \epsilon -\eta -\epsilon_{Q} \right) + \, \epsilon\,\rho_{Q}\\
\frac{\text{d} \rho_{Q}}{\text{d} N}= \rho_{Q} \left(\theta - 2\eta -\epsilon_{Q}\right) + {}^{2}\chi_{Q}\\
\vdots 
\\
\frac{\text{d}  {}^{l}\chi_{Q}}{\text{d} N}= {}^{l}\chi_{Q} \left[l\theta -\left(l-1\right)\epsilon - \left(l+1\right)\eta -\epsilon_{Q} \right] + {}^{l+1}\chi_{Q}
\end{cases}  
\label{eq:system_Q}
\end{equation}
Solving the system we can predict the evolution of any generic quantity $Q(\phi)$ that will depend also on the background via the slow-roll parameters $\epsilon$, $\eta$ and $\theta$, as expected. This means that, in principle, one can evolve all the $M$ coefficients in Eq.~\eqref{eq:DeltaS} and study different models of inflation in full generality.

\subsection{Theoretical Monte Carlo: Integration scheme}

Our aim is to explore a reasonably large ensemble of physical models of inflation that can lead to a sizable gravitational wave production and calculate their contribution to the energy-density of the early Universe, accurately. In this regard, it is worth noting that taking into account all the operators in the quadratic effective action that induce tensor perturbations, one can derive the following leading order relation for the power-spectrum~\citep{Creminelli:2014wna,Noumi:2014zqa,Giare:2020vss}:
\begin{equation}
  \mathcal P_{\rm T}= \frac{1}{c_{\rm T}}\left(\frac{H^2}{\pi^2\,M_{\mathrm{pl}}^{2}} \right)
  \label{eq:PtCt}
\end{equation}
where $c_{\rm T}$ is the propagating speed of tensor modes that can be simply expressed in terms of $\bar{M}_3$ as $c_{\rm T}^{-2}=1 - \bar{M}_3^2 / M_{\mathrm{pl}}^{2}$ where $\bar{M}_3$ is defined in~\eqref{eq:EFT_action}. In this case, it is straightforward to see, from its definition, that the tensor tilt acquires a further correction
\begin{equation}
    n_{\rm T}=-2\epsilon + \epsilon_{\rm T}
    \label{ntt}
\end{equation}
where the evolution of the parameter
\begin{equation}
    \epsilon_{\rm T}= - \frac{\dot{c}_{\rm T}}{H\,c_{\rm T}}
    \label{ctepsilont}
\end{equation}
is clearly governed by the system \eqref{eq:system_Q}. It is also worth noting that in this framework the standard relation between the tensor amplitude and the tensor tilt does not hold anymore and more general consistency relations can be derived both in the absence and in presence of additional EFT operators, see  Refs.~\citep{Giare:2020vss, Capurri:2020qgz} for detailed discussions. Anyway, all the cosmological observables can still be expressed in terms of the slow-roll parameters and in particular, the tensor spectrum and its evolution are fully determined by the evolution of the background and the parameter $\epsilon_{\rm T}$. This is an important achievement since through the flow equation method we can actually test the observable predictions of a large number of stochastically generated models, without relying on the specific form of their underlying actions. To optimize our model exploration, we proceed with a theoretical Monte Carlo as follows:

\begin{itemize} [leftmargin = *]

\item First and foremost, we notice that the hierarchy of flow equations must be truncated at finite order, which we choose to be the $4$th-order. Then, we draw a suitable set of randomly chosen initial conditions for the background parameters.  In particular, we randomly choose the parameters introduced in the Hubble-tower \eqref{Htower} within the following ranges
\begin{gather*}
    \epsilon_{\rm in}\in[0,0.8],\\
    \eta_{\rm in}\in[-0.1,0.1],\\
    ^2\lambda_{\rm in}\in[-0.05,0.05],\\
    ^3\lambda_{\rm in}\in[-0.005,0.005],
\end{gather*}
while for the $c(\phi)$-tower \eqref{Ctower} the initial conditions are taken from the sets
\begin{gather*}
    \theta_{\rm in}\in[-0.1,0.1],\\
    \kappa_{\rm in}\in[-0.1,0.1],\\
    ^2\xi_{\rm in}\in[-0.05,0.05],\\
    ^3\xi_{\rm in}\in[-0.005,0.005].
\end{gather*}
These ranges are very similar to those in \citep{Kinney:2002qn} and \citep{Capurri:2020qgz}.

\item Once the initial conditions are chosen, we solve the Hubble flow equations \eqref{eq:HFE_background} for the background slow-roll parameters. Specifically, we integrate the equations forward in time for at most $\sim10^4$ e-folds of expansion. Then, apart from the unfortunate cases where the integration did not survive, we expect two possible outcomes: either we reach a fixed point (that we eliminate) or we manage to get the end of inflation defined by the usual relation $\epsilon=1$. In this latter case, we store all the background parameters as functions of the number of e-folds $N$ before the end of inflation (\textit{i.e.}, $N=0$ corresponding to $\epsilon=1$). Given a large number of repetitions ($\gtrsim 10^4$), approximately $90\%$ of the time the end of inflation is successfully reached. 

\item We then check that the models stored in the previous point allow a sufficient long phase of expansion and are able to explain observations. To do so we use the values reached by parameters at the end of inflation as new initial conditions at $N=0$ and perform a backward-in-time integration up to the e-folds when the primordial observables are evaluated ($N=60$). Once more, we make sure to obtain a successful integration, that is, we do not end up with $\epsilon=1$ again. For the remaining models, we check whether the spectral index of scalar modes $n_s$ lies within the observed bounds. In particular, we reject all the results outside the range $0.94<n_s<0.98$, chosen conservatively around the Planck best-fit value, ending up with roughly $17\%$ of the total. We store the survived models and proceed to evolve all the other physical quantities involved in our analysis.   

\item  Particularly relevant for the evolution of the tensor spectrum are the quantities related to the propagating speed $c_T$, see also Eq.~\eqref{eq:PtCt}. To solve the system of equations~\eqref{eq:system_Q} we need to specify some initial conditions for $\epsilon_T$ and the other high-order tensor parameters that  we randomly choose within the following ranges:
\begin{gather*}
    \epsilon_{\rm T}\in[-0.1,0.1],\\
    \rho_{\rm T}\in[-0.01,0.01],\\
    ^2\chi_{\rm T}\in[-0.001,0.001],\\
    ^3\chi_{\rm T}\in[-0.0001,0.0001].
\end{gather*} 
To optimize the simulations and save computational time,  for each realization of the background, our algorithm is able to perform simultaneous evolution of different physical quantities. In particular, starting from some initial conditions, we first perform a forward integration until the end of Inflation. Since we already did such an integration for the background (and given that all the other quantities do not affect the spacetime evolution), we can focus exclusively on the stability of the tensor-speed sector. We find that a small part of the total leads to an unsuccessful integration while most models require also a backward integration (for instance because they reach the controversial value $\epsilon_T=1$ during the integration or because they show non-physical behaviors for the other parameters). Once that all the consistency checks have been carried out, the model is either accepted or rejected. At the end of the process, only approximately $40\%$ of the attempts resolve in a successful inflation with a non-trivial tensor-speed sector.

\item As concerns the other physical quantities, we select a subgroup of models that share the same tensor amplitude $r\sim 0.001$ on the CMB scales ($N=60$) but that differ by the value of the tensor tilt $n_T$ that we estimate at $N=60$ according to Eq.~\eqref{ntt}. Finally, we evolve the tensor spectrum dynamically from $N=60$ up to the end of inflation by means of the Hubble flow Equations. For each spectrum, we calculate the corresponding contribution to the energy density of the early universe parametrized in terms of the effective number of relativistic degrees of freedom $\Delta N_{\rm eff}^{\rm GW}$. To do so, we evaluate the corresponding energy-density in gravitational waves $\Omega_{\rm GW}(f)$ by Eq.~\eqref{OmegaGW} and integrate it over frequency according to  Eq.~\eqref{Int1}.

\end{itemize}

\subsection{Theoretical Monte Carlo: Models}
Using this procedure, we are able to collect a sufficiently large ensemble of physical models ($\simeq 10.000$) which spans a reasonable range of possibilities, from realization with a canonical tensor-speed sector (\textit{i.e.}, $c_{\rm T}=1$  and $\epsilon_T=0$) to more general cases with time-dependent tensor parameters. Nonetheless, our integration scheme is focused on a well-defined task, \textit{i.e.}, we are not interested in providing a comprehensive analysis of the model frequency distribution for the different observables as already done in full generality in Ref.~\citep{Capurri:2020qgz}, but rather to shed light on the correlation between $n_{\rm T}$ and the predictions for $\Delta N_{\rm eff}^{\rm GW}$. To achieve this task in the most direct and simple way we necessarily introduce some limitations on the models that we are actually able to explore, that deserve to be further justified and clarified.

A first major restriction comes from limiting our analysis to a small subgroup of models with a fixed tensor amplitude $r\sim 0.001$ on the CMB scales. This clearly introduces a limitation on the number of cases that we are able to reach within our Monte Carlo technique. Notice however that we do not expect this limitation to introduce a large bias on the frequency distribution of the values obtained for the tensor tilt as the consistency relation between $r$ and $n_T$ does not hold anymore and these two parameters can be regarded as independent. In addition, we are not particularly interested in studying the model frequency distribution, but rather in understanding whether models sharing similar parameters on the CMB scales may result into a significant different contribution to the energy budget of the early Universe because of their different evolutionary paths. Focusing only on models with the same $r$ at $N=60$ turns out to be particularly useful for this purpose since it ensures that the predictions for $\Delta N_{\rm eff}^{\rm GW}$ do not depend on the value of the tensor amplitude at the CMB scales (which is in fact common to all models). In this way at $N=60$ all the models will differ only by the value of the tensor tilt and comparing the values of $\Delta N_{\rm eff}^{\rm GW}$ predicted by models with a similar $n_{\rm T}$ one can have an immediate idea of the difference produced by the different evolution of the spectra from $N=60$ to $N=0$ and unequivocally understand whether $\Delta N_{\rm eff}^{\rm GW}$ and $n_{\rm T}$ are somehow correlated. Finally, we can directly compare the results obtained within our theoretical Monte Carlo with those derived in \autoref{sec.3.3} by means of a parametric reconstruction of the spectra (where the tensor amplitude was fixed to $r\sim 0.001$, as well), testing the consistency of these two methods. Last, but not least, $r\sim 0.001$ is the declared target of future CMB-S4-like experiments~\citep{CMB-S4}. Therefore we believe it should be particularly interesting to understand what kind of physical models future surveys may be able to probe. This is the ultimate reason why we have chosen such a value for the amplitude. 

A second minor limitation is introduced by taking only positive initial conditions for the parameter $\epsilon$, without considering models resulting from a background evolution with $\epsilon_{\rm in}<0$, like it was done in Ref.~\citep{Capurri:2020qgz}. To understand the implications of this limitation, we recall that in the standard single-field models the Null Energy Condition (NEC) prevents the slow-roll parameter $\epsilon$ to be negative. However, this framework is quite general and can be applied also to more complicated scenarios where this possibility is viable, such as super-inflation models~\citep{Creminelli:2006xe,Gasperini:1992pa,Brustein:1995ah,Baldi:2005gk} (where $\epsilon$ can remain always negative) or models with intermittent NEC violation~\citep{Cai:2020qpu,Cai:2022nqv} (where $\epsilon$ can be negative for some e-folds and then come back to be positive, restoring the usual end of inflation at $\epsilon=1$). In this regard, we notice that starting with a positive $\epsilon$ as the initial condition does not preclude this parameter to acquire negative values during its evolution. Therefore the latter intermittent case is included in our Monte Carlo. Conversely, requiring $\epsilon_{\rm in}>0$ and the end of inflation to occur at $\epsilon=1$ exclude the super-inflation case. Indeed such models are characterized by an Hubble parameter that increases with time so that the end of inflation is no longer determined by the condition $\epsilon=1$ but must be forced by external factors, such as an additional field. In the framework of a theoretical Monte Carlo, it becomes very ambiguous to decide when inflation ends since we are not sensitive to the details of the mechanism. For this reason, one needs to choose an arbitrary point during the evolution, as done in Ref.~\citep{Capurri:2020qgz}, introducing an element of arbitrariness in the resulting predictions. In addition, this case should be considered separately because one needs to choose the range of integration very carefully in such a way that the energy scale at the end of inflation can lie whithin the observational upper bound ($H_{\rm fin} < 2.7 \times 10^{-5} \,\text{M}_{\rm pl}$~\citep{Akrami:2018odb}) and the lower limit (around the MeV scale to guarantees hydrogen and helium production during the BBN~\citep{Giudice:2000ex}). 
It is important to acknowledge that this limitation can in fact result in a significant reduction in the number of models predicting a blue-shifted tensor tilt that our pipeline is able to investigate, see also \autoref{fig:figure3}. Despite this, our conclusions on $\Delta N_{\rm eff}^{\rm GW}$ cannot in any way rely on these exotic scenarios and we can safely exclude such models from the analysis without biasing the results. For a more thorough examination, see Ref.~\citep{Capurri:2020qgz}.

\subsection{Observable Predictions: Primordial Tensor Spectrum}

\begin{figure}
 	 \centering
 	 \includegraphics[width=\columnwidth]{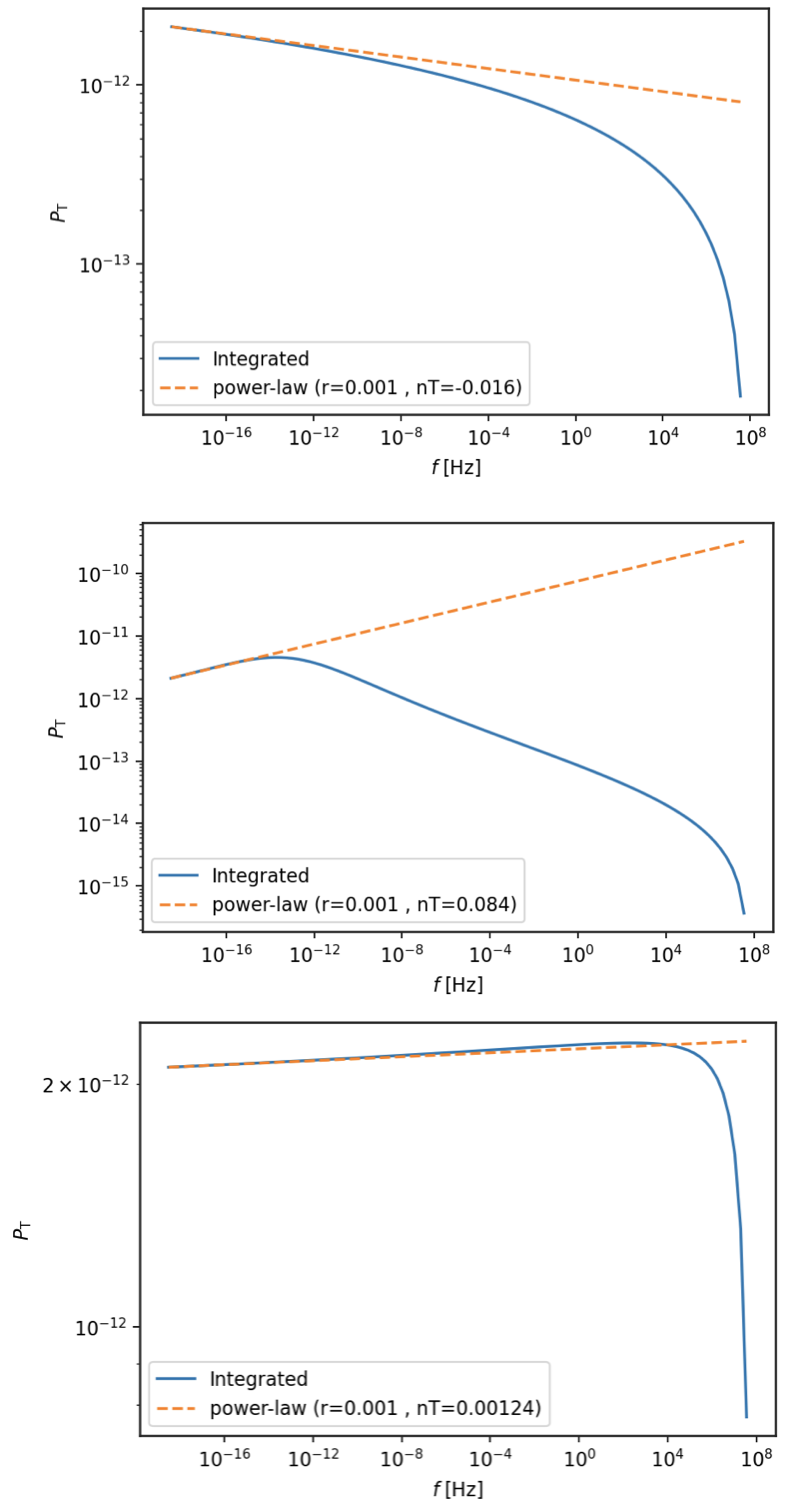}
 	 \caption{Three physical examples of the primordial tensor spectra (and their power-law extrapolation, orange line) obtained by integrating the Hubble Flow Equations as discussed in \autoref{sec.4}.}
 	 \label{fig:figure4}
\end{figure}

We start discussing some useful insights about what is obtained following the integration method outlined in the previous subsection. In particular, in \autoref{fig:figure4} we show three controversial examples that we find particularly enlightening about the very diversified behavior that the tensor spectrum can acquire in physical models of inflation, validating the need to systematically investigate the observable predictions by means of a theoretical Monte Carlo. 

In the top-side panel of the \autoref{fig:figure4}, we display the spectrum predicted by a red-tilted model of inflation where we evolved the background according to Eq.~\eqref{eq:HFE_background}, switching off all the other EFT operators. The shape of this spectrum is not very different from what we got by our previous parametric analysis (see also \autoref{fig:figure2}). Not surprisingly, near the CMB frequencies, the spectrum is well described by a power-law (orange line in the figure) and all the physics of the model is captured only by two parameters, the amplitude and the tilt. Conversely, on frequencies close to the end of inflation the power in gravitational waves is suddenly dismissed. As soon as the potential start approaching its minimum (\textit{i.e.}, $\epsilon \to 1$) the slow-roll dynamics breaks down and both the Hubble parameter and the tensor spectrum ($\mathcal P \propto H^2$) suddenly decrease. In this frequency range, the behavior of the spectrum is mostly determined by the shape of the potential (which is no more flat) and consequently, the gravitational wave production becomes strongly model-dependent~\citep{Kinney:2021nje}. Interestingly, if we compare the power spectrum integrated over the Hubble flow equations with a simple power-law extrapolation, we see that on high frequency there is a difference of almost two orders of magnitude between the two curves. We can easily quantify the impact in terms of $\Delta N_{\rm eff}^{\rm GW}$ by integrating both the spectrum and its power-law extrapolation through Eq.~\eqref{Int1}. For this particular model (and for models that show a similar behavior) we estimate a difference of a factor $\sim 10$ between the contributions obtained by integrating the Hubble flow equation ($\Delta N_{\rm eff}^{\rm GW}\simeq 1 \times 10^{-12}$) and the one inferred by a power-law extrapolation ($\Delta N_{\rm eff}^{\rm GW} \simeq 1 \times 10^{-11}$). In both cases, however, the contribution is extremely small and well beyond any current or future experimental sensitivity, as expected in red-tilted inflation.  

The situation becomes even more intriguing if we turn to the study of blue-titled models of inflation. Within our framework, such models can be realized either taking $\epsilon<0$ at $N\sim 60$ or including corrections to the tensor spectrum coming from the extrinsic curvature perturbations in Eq.~\eqref{eq:EFT_action}. In the middle panel of \autoref{fig:figure4} we plot the tensor spectrum realized in one of the latter cases. In this particular model, the regime $n_{\rm T}>0$ is supported only for a few e-folds of inflation, corresponding to the frequency range during which the spectrum follows a blue-tilted power-law and the gravitational signal is amplified. After that, because of a combined effect of the background evolution and the evolution of parameter $\epsilon_{\rm T}$, the spectrum becomes very red-tilted and the power in the gravitational wave is suppressed at high frequencies. Specifically, the more $n_{\rm T}$ is positive at CMB scales, the greater $\epsilon_{\rm T}$ should be, bringing consequently its derivatives to assume larger values to compensate. Thus, its evolution is accelerated (towards negative values). The blue-tilted regime lasts only a few e-folds and then fall into the red-shifted one.  This model is similar to those discussed in Ref.~\citep{Benetti:2021uea} and in this case assuming a blue-tilted power-law spectrum over all scales leads to overestimating the gravitational wave signal by a factor of $10^{5}$. Repeating the exercise of computing the contribution to the radiation energy-density for both the integrated spectrum ($\Delta N_{\rm eff}^{\rm GW} \simeq 7 \times 10^{-14}$) and the power-law one ($\Delta N_{\rm eff}^{\rm GW} \simeq 2 \times 10^{-9}$) we end up with two completely different results. Therefore this is the "smoking gun" evidence that leads weight to all the concerns already emerged from our parametric analysis. It makes evident that extrapolating a power low spectrum over all scales can be an unreliable practice and can lead to strongly overestimating the gravitational wave contribution to the radiation energy-density (even by many orders of magnitude, as we have just proved). However one may ask to what extent such a model can be considered representative of the spectrum's behavior in blue-tilted inflation and how easily models like this one can be realized. As a counterexample, in the bottom panel of \autoref{fig:figure4} we show a blue-tilted spectrum realized within a model where a simple power-law extrapolation still provides a very good approximation of the gravitational wave production even on a frequency very close to the end of inflation, guarantying an accurate estimation of $\Delta N_{\rm eff}$. Notice that models like that are typically characterized by an extremely slow evolution of the inflationary parameters and hence by a very flat potential. Therefore one may argue that they may be not easy to realize, as well. 

\begin{figure}
 	 \centering
 	 \includegraphics[width=\columnwidth]{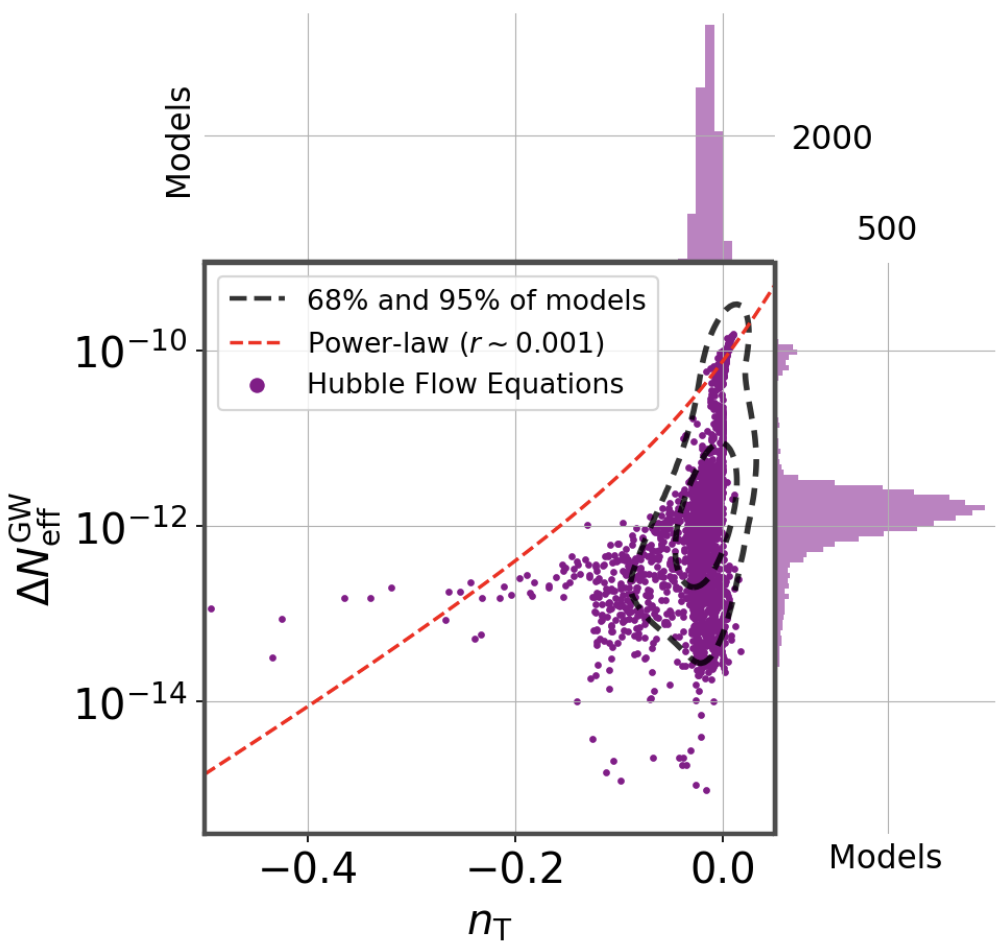}
 	 \caption{\small Observable predictions in the plane $(n_{\rm T}\,,\,\Delta N_{\rm eff}^{\rm GW})$. The magenta dots represent the models realized within the Hubble Flow Equation method discussed in \autoref{sec.4} while the red dashed line represents the power-law prediction. The dashed black lines define the regions of the plane that contain the 68\% and 95\% of the total models and are calculated by marginalizing over the point frequency distribution (displayed by the two histograms on the axes).  }
 	 \label{fig:figure5}
\end{figure}

Clearly to provide a definitive answer and derive reliable results we need to study the inflationary gravitational wave production for a sufficiently large ensemble of randomly realized physical models where all the possibilities are studied exhaustively in the framework of a Theoretical Monte Carlo.  In \autoref{fig:figure3} we compare the results for $\Delta N_{\rm eff}^{\rm GW}$ obtained by this latter approach (dark magenta dots in the figure) with those realized in \autoref{sec.3.3}. In \autoref{fig:figure5} we instead zoom-in on the $(n_{\rm T}\,,\, \Delta N_{\rm eff})$ plane, showing the distribution of the physical models. The dashed black lines in this latter figure define the regions of the plane where the 68\% and 95\% of the total models lie and are calculated by marginalizing over the point frequency distribution (displayed by the two histograms on the axes). An accurate analysis of this figure can reveal several interesting hints about the physics underlying the point distribution that is worthy of being discussed in details. 

We start by analyzing the observable predictions for the tensor tilt. In particular, we notice that the vast majority of the models are characterized by slightly negative tilt ($n_{\rm T}\lesssim 0$). As partially explained in the previous subsection, this is due to the fact that, during the integration process, only a few blue-tilted models survive all the physical consistency checks and constraints. In fact, most of the survived models have a canonical tensor speed evolution ($c_{\rm T}=1$, $\epsilon_{\rm T}=0$) and respect the null energy condition ($\epsilon>0$) so that their observable predictions follow, or are very close to following, the usual slow-roll consistency relations. This also suggests that realizing well-defined blue-tilted models able to satisfy all the physical requirements (such as stability, causality and last but not least the observable constraints) may be a tricky avenue and in general red-tilted models are largely preferred by theoretical Monte Carlo simulations, as already pointed out in Ref.~\citep{Capurri:2020qgz}. Focusing on the survived blue-tilted models, it is also evident that only small values of the tensor tilt are realized and we remain far away from the controversial observational upper limit inferred for this parameter. As a matter of fact, the largest $n_{\rm T}$ we are able to get within our pipeline reads $n_{\rm T}\simeq 0.08$ (close to the middle panel of \autoref{fig:figure4}). The reasons why the case $n_{\rm T}>0$ is generally disfavored are several. For instance, such values can hardly arise from the extrinsic curvature corrections since this would imply a large $\epsilon_{\rm T}>0$ and, by Eq.~\eqref{ctepsilont}, a negative time derivative of the tensor speed that would thus be reduced close to the frequencies where it is instead constrained to be unitary by gravitational wave observations~\citep{GBM:2017lvd,Monitor:2017mdv}. As concerns the red-titled models, most of them show the same preference for very small tilt values ($-0.1<n_{\rm T}<0$  within the 95\% region), but a few exceptions with $n_{\rm T}\lesssim -0.2$ can be observed. While they represent a negligible part of the total points, it is interesting to notice that, in principle, such models do not violate any observable prediction. In our framework, a relatively large negative tilt can be realized by a combined effect of both the background evolution and the tensor speed evolution provided that $\epsilon_{\rm T}<0$. In this regard, it is worth pointing out that a negative $\epsilon_{\rm T}$ would imply $\dot{c}_{\rm T}>0$ and so a tensor speed that increases over time, around the CMB frequencies. Since the propagating speed of gravitational interactions is not (severely) constrained at those frequencies, the model may remain viable as long as $c_{\rm T}\in [0\,,\,1]$, see also Refs.\citep{Capurri:2020qgz,Giare:2020vss}. On the other hand, a significantly non-unitary $c_{\rm T}$ may be an element of concern because we are dealing with perturbative departures from General Relativity and so we do not expect large deviations. However, thanks to the narrowed window allowed for the initial conditions of the tensor-speed parameters, these models remain the very minority (we can count only 18 models with $n_{\rm  T}\lesssim -0.2$) and in most of them, the background dynamics importantly contributes to this behavior. We, therefore, find this issue to be not particularly relevant to the general aim of this paper and leave it suitable for future investigation. There is yet another interesting aspect that deserves to be remarked: looking at \autoref{fig:figure5} we can spot a bunch of models that follow a power-law behavior very closely (dark magenta points that overlap with the red dashed line in the figure). All these models are characterized by a tensor tilt extremely close to zero. Because of Eq.~\eqref{ntt}, this means that both $\epsilon$ and eventually $\epsilon_{\rm T}$ need to be very close to vanishing, implying an extremely slow-roll dynamics and thus a very flat inflationary potential. They are nothing but models that behave like the one depicted in the bottom panel of \autoref{fig:figure4}. So, looking at \autoref{fig:figure5}, we can finally answer whether such models can be easily realized or not. In particular, we find that they are not in the densest region of the plane. Nonetheless, they still fall within the region containing the 95\% of the total models, actually contributing to a second (very) small peak in the histograms of $\Delta N_{\rm eff}$, see also \autoref{fig:figure5}.

\subsection{Observable Predictions: Relic Gravitational Radiation}
We now turn to the study of the observable predictions for $\Delta N_{\rm eff}$ that is the point of interest for this analysis. First and foremost, most of the models predict a value very different from what one would expect by extending a power-law parameterization over all scales, see \autoref{fig:figure3}. The same conclusion can be derived from a different perspective by looking at \autoref{fig:figure5}. From the latter figure we can appreciate that, while the histogram of $n_{\rm T}$ is very sharped and most models share similar values of the tensor tilt, the histogram of $\Delta N_{\rm eff}^{\rm GW}$ is instead much broader and the regions containing the 68\% and 95\% of models are almost vertical, spanning a quite large range of values $\Delta N_{\rm eff}^{\rm GW}\simeq 10^{-10}\, - \, \simeq 10^{-14}$. This means that models that share the same inflationary parameters on the CMB scales (\textit{i.e.}, the same amplitude $r$ and the same tilt $n_{\rm T}$) can easily result in a completely different contribution to $\Delta N_{\rm eff}^{\rm GW}$. As already pointed out in \autoref{sec.3}, this depends on the different evolution of the spectra at high frequencies. More precisely, the results of theoretical Monte Carlo suggest that extrapolating a power-law behavior over ultraviolet frequencies, in most cases leads to overestimating the gravitational wave contribution to the energy budget of the Universe. As partially explained above, this is due to the fact that, close to the end of inflation, the potential necessarily undergoes a rapid phase of evolution towards its minimum that drives the Hubble parameter (and consequently the power spectrum) to be suddenly dismissed. This is evident in all the three physical spectra shown in \autoref{fig:figure4}. This feature is instead missed within a power-law extrapolation so that the power in gravitational waves is typically much overestimated on high frequencies, leading to a larger $\Delta N_{\rm eff}^{\rm GW}$. Nonetheless, we can observe a few models where the actual contribution to the effective number of relativistic species is larger than predicted by a power-law. While such points represent the vast minority of the models, it is still interesting to explain the physical reason underlyng this behavior. In particular, it is evident both from \autoref{fig:figure3} and from \autoref{fig:figure5} that this event is more frequent for those very few points that show a very red tensor tilt $n_{\rm T}\lesssim - 0.2$. As we pointed out in the previous paragraph, in these cases both the background dynamics (\textit{i.e.}, the value of $\epsilon$ at $N=60$) and the tensor-speed dynamics (\textit{i.e.}, the value of $\epsilon_{\rm T}$ at $N=60$) should significantly contribute to the final value of the tensor tilt on the CMB scales and to the evolution of the spectrum with frequency. Indeed, in all these models, $\epsilon$ must reach the value $\epsilon=1$ within $\Delta N=60$ e-folds of evolution (so that inflation can end) while $\epsilon_{\rm T}$ will undergo a similar evolution. If $\epsilon_{\rm T}$ evolves towards less negative values, it can mitigate the loss of power induced by $\epsilon$ and the spectrum may remain so much red-tilted only for a few e-folds. Consequently, in this case assuming continuously a power-law parameterization can lead to underestimating $\Delta N_{\rm eff}^{\rm GW}$.

We conclude this section with a final remark: all the models obtained with the Hubble Flow Equation method give an extremely small $\Delta N_{\rm eff}^{\rm GW}$. The histogram of this parameter is in fact centered around values $\Delta N_{\rm eff}^{\rm GW}\sim 10^{-12}$, with a second small peak of models at $\Delta_{\rm eff}^{\rm GW}\sim 10^{-10}$ (resulting from that class of models with an extremely slow evolution discussed in the previous paragraphs). These values are far away from the total amount of additional radiation allowed by data ($\Delta N_{\rm eff}^{\rm GW}\lesssim 0.3 - 0.4$) as well as from any current and future experimental sensitivity. Therefore one may ask whether this issue is in any way relevant for the purpose of mode-building. In this regard, we would like to point out that, while this method is quite general and allow precise calculation without relying on the details of the model, it does not cover all the possibilities proposed in the literature and blue-tilted models with larger gravitational wave production may be obtained by other viable physical mechanisms. On the other hand, the stochastic technique used in \autoref{sec.3.3} should embrace a much larger class of possibilities since we simply reconstruct the spectrum as a fraction of the frequency. It is also entirely plausible that such spectra can be obtained in well-motivated models, as we may argue by comparing the gray and dark magenta dots in \autoref{fig:figure3} and noticing that they share similar behavior. In any case a detailed study of the observational prospects of the field is beyond the aim of this manuscript where we believe to have already covered a reasonable range of different scenarios and possibilities, consistently getting conclusive evidence that assuming a power-law spectrum over all scales can lead to a wrong estimation of the gravitational wave contribution in the early Universe. In light of this result, we can definitively conclude that the calculation of $\Delta N_{\rm eff}^{\rm GW}$ proves to be remarkably model-dependent and more accurate analyses are needed before inferring any reliable conclusion on (blue-titled) inflationary models in light of the BBN bounds on additional radiation. 

\section{Conclusions}
\label{sec.5}
In this paper we revisit the calculation of the inflationary gravitational wave contribution to the radiation energy-density in the early Universe. Behaving as additional radiation, primordial gravitational waves may in fact increase the effective number of relativistic species ($N_{\rm eff}$) by a further correction that depends on the integrated energy-density in gravitational radiation over all scales, see Eq.~\eqref{Int1}. According to the Friedmann equations, extra radiation would imply a faster background expansion and consequently a different thermal evolution of the Universe, with several implications. For instance, a faster expansion would lead to a higher freeze-out temperature of the weak interactions, implying a higher fraction of primordial Helium and Deuterium to be forged during the Big Bang Nucleosynthesis epoch. This effect is particularly relevant, because it is commonly used to infer stringent bounds on the additional radiation energy-density and, in its turn, to constrain (blue-titled) models of inflation. 

However the underlying assumption of (most of) the state-of-the-art analyses is that the spectrum of inflationary gravitational waves can be parametrized, \textit{continuously} over all cosmological epochs and scales, by a simple power-law with two free parameters: the amplitude $r$ and the tilt $n_{\rm T}$. While in most inflationary models such parameterization works very well on the frequencies probed by the CMB experiments (roughly corresponding to $N\sim 60$ e-folds before the end of inflation), as already pointed out in the literature~\citep{Giare:2020vhn,Kinney:2021nje} extrapolating a power-law behavior over all frequencies can be highly non-trivial and risky; above all on the high-frequencies corresponding to tensor modes that cross the horizon very close to the end of inflation, when the slow-roll dynamics breaks down and the gravitational wave production becomes strongly model-dependent. Since these frequencies not only contribute to the integral~\eqref{Int1}, but they are also exponentially amplified within a power-law parameterization ($\mathcal P(f)\propto f^{n_{\rm T}}$), this problem becomes of primary relevance when evaluating the tensor modes contribution in the early Universe because the calculation crucially depends on a parameterization whose validity is anything but reliable.  

Driven by this concern, in \autoref{sec.3}, we systematically study how (much) different parameterizations of the tensor spectrum
impact on the final predictions of $\Delta N_{\rm eff}^{\rm GW}$. In \autoref{fig:figure1} we show that allowing a $\sim \text{few} \%$ scale-variation of the tensor tilt, the resulting $\Delta N_{\rm eff}^{\rm GW}$ can be much amplified or suppress, depending on the sign of the running. In \autoref{sec.3.3} we perform parametric analysis by expanding the spectrum in full generality as a sum of powers and randomly collecting $10^{6}$ different shapes of the spectrum able to satisfy all the observational constraints, consistently towards all cosmological epochs and scales. The results in \autoref{fig:figure3} (gray dots) prove that relaxing that assumption of power-law spectrum \textit{on high frequencies}, the value of the tensor tilt becomes basically uncorrelated with $\Delta N_{\rm eff}^{\rm GW}$ so that models with the same $n_{\rm T}$ can contribute very differently to the energy budget of the Universe.

In order to understand to what extent our result can be considered reliable when applied to physical models of Inflation, in \autoref{sec.4} we investigated the observable predictions of a very broad class of inflationary models. We work within the framework of the Effective Field Theory of inflation and follow a methodology based on the so-called Hubble Flow Equation: a system of coupled differential equations whose solution completely specifies the evolution of the main observable quantities during inflation. We solve numerically the truncated system of the Hubble Flow Equations for a set of suitably defined initial conditions (taking into account also a different combination of additional operators in the EFT of inflation) as a sophisticated algorithm for generating large numbers of slow-roll inflationary models without relying on the explicit form of the action. In this way, we produce an ensemble of very general physical models ($\simeq 10.000$) studying the resulting observable predictions. Examples of the spectra obtained by our method are shown in \autoref{fig:figure4} while the final results for $\Delta N_{\rm eff}^{\rm GW}$ are summarized both in \autoref{fig:figure3} and in \autoref{fig:figure5}. Both figures make it evident that in most cases extrapolating a power-law behavior over 24 orders of magnitude in frequency leads to overestimating the power in gravitational waves, above all on the ultraviolet frequencies that are the most relevant in the calculation. As a result, the predicted relic energy-density in gravitational wave can be ultimately incorrect. 

We conclude by stressing that this issue seriously calls into question the validity of the observational constraints inferred on the tensor tilt by the indirect effect of additional radiation during the BBN epoch, motivating the need of more accurate calculations before inferring any reliable conclusion on inflation.

\section*{Acknowledgements}
WG thanks Martina Gerbino and Massimiliano Lattanzi for the useful comments at the early stage of this project and Fabrizio Renzi for valid suggestions on the manuscript. WG, MF and AM are supported by "Theoretical Astroparticle Physics" (TAsP), iniziativa specifica INFN. EDV is supported by a Royal Society Dorothy Hodgkin Research Fellowship.
\section*{Data availability}
The data underlying this article will be shared on reasonable request to the corresponding author.
\bibliographystyle{mnras}
\bibliography{MNRAS} 

\appendix

\section{Gravitational Radiation in the early Universe}
\label{Appendix-A}

We consider a spatially flat FLRW metric, whose perturbed line element
in synchronous gauge reads~\citep{Ma:1995ey}
\begin{equation}
d s^{2}=a^{2}(\eta)\left[d \eta^{2}-\left(\delta_{i j}+h_{i j}\right) d x^{i} d x^{j}\right]
\end{equation}
with $a$ and $\eta$ denoting scale factor and conformal time, respectively. In this picture, generic tensor perturbations are described by the transverse and traceless part of the symmetric $3 \times 3$  matrix $h_{ij}$. In the Fourier space, focusing on one particular polarization state and a given mode $k$, the gravitational weave field satisfies the usual equation of motion\footnote{It should be noted that, as is commonly done in the literature, Eq.\eqref{eq:motion} does not take into account the damping of primordial gravitational waves caused by an anisotropic stress tensor. This is relevant for free-streaming thermal neutrinos and other high-energy particles, see, \textit{e.g.}, Refs~\citep{Weinberg:2003ur,Watanabe:2006qe,Benini:2010zz,Dent:2013asa}. Although these corrections are typically small (around 30\%), the damping can potentially affect the power-law behavior of the tensor spectrum. However, incorporating these effects into the current analysis is not straightforward and beyond the scope of this work. This issue is left suitable for future studies.}~\citep{Lyth:2009}
\begin{equation}
h_{k}^{\prime \prime}+2 \mathcal{H} h_{k}^{\prime}+k^{2} h_{k}=0
\label{eq:motion}
\end{equation}
where the prime indicates the derivative with respect to the conformal time and $\mathcal H = a^{\prime}/a$. Since here we are mainly interested in primordial gravitational waves, it is particularly convenient to characterize the gravitational field in terms of its power spectrum 
\begin{equation}
\mathcal P_{\rm T}(k) = \frac{2 k^3}{\pi} \, \left |h_k(\eta_i)\right|^2 
\end{equation}
where, for each mode $k$, $h_k(\eta_i)$ specifies the value of the field at some initial conformal time $\eta_i$. In this way, connecting this picture to inflation simply requires identifying the power spectrum of the gravitational field with the primordial spectrum of inflationary tensor modes. 

In the early Universe, a satiable background of gravitational waves will clearly increase the energy-budget by providing an additional form of radiation. Here we parameterize this contribution in terms of corrections to the effective number of relativistic degrees of freedom $N_{\rm eff}$. Within the Standard Model of particle physics this parameter acquires the reference value of $N_{\rm eff} = 3.044$~\citep{Mangano:2005cc,2016JCAP...07..051D,Akita:2020szl,Froustey:2020mcq,Bennett:2020zkv}, counting three different families of relativistic neutrinos plus an additional contribution coming from the non-instantaneous neutrino decoupling. To understand how this reference value is modified in presence of additional gravitational radiation, we focus on temperatures $T\gtrsim \mathcal O(1)$ MeV when the relativistic species in the Universe were electrons (and their antiparticles, positrons) $e^{\pm}$, neutrinos $\nu$ and photons $\gamma$. Including also the contributions of gravitons, the total amount of radiation will read \citep{Maggiore:1999vm}
\begin{equation}
\rho_{\rm rad} = \frac{\pi^2}{30} \left[ 2 \, T_{\gamma}^4 + \frac{7}{4}\, T_{e^{\pm}}^4 + \frac{7}{4} \,N_{\rm eff}\, T_{\nu}^4 + 2 \, T_{\rm GW}^4 \right]
\end{equation}
where the factor 2 in front of $T_{\rm GW}$ counts the two different polarization states $(+,\times)$ of tensor perturbations. Apart from the gravitons, all the other species were in thermal equilibrium and shared the same temperature: $T_{\gamma}=T_{e^{\pm}}=T_{\nu}$. Therefore it is straightforward to see that we can describe gravitational radiation as an  additional contribution to the effective number of relativistic species
\begin{equation}
\Delta N_{\rm eff}^{\rm GW} =  \frac {8}{7} \frac{T_{\rm GW}^4}{T_{\gamma}^4}=\left. \frac {8}{7} \frac{\rho_{\rm GW}}{\rho_{\gamma}}\right|_{\rm T_{\gamma}\gtrsim \mathcal O(1) \,\text{MeV}}
\end{equation}
To re-scale this contribution to the present time, we must consider that after $T\gtrsim \mathcal O(1)\,\rm{MeV}$, as the the Universe expands, the gravitational wave energy-density decays as $\rho_{\rm GW}\sim 1/a^4$, while, assuming entropy conservation, the CMB photon energy-density evolves as $\rho_{\gamma} \sim 1 /\left(a^{4} g_{*, s}^{4 / 3}\right)$ with $g_{*, s}$ the number of entropic degrees of freedom. Therefore the present-day contribution will be given by
\begin{align}
\Delta N_{\rm eff}^{\rm GW} &=\left.\frac{8}{7}\left(\frac{g_{*, s}(T \gtrsim 1 \mathrm{MeV})}{g_{*, s}\left(T_{0}\right)}\right)^{\frac{4}{3}} \frac{\rho_{\mathrm{GW}}}{\rho_{\gamma}}\right|_{\rm Today}
\end{align}
with $g_{*, s}(T_0)\simeq 3.91$ the current number of entropic degrees of freedom. 

While the present Cosmic Microwave Background energy density $\rho_{\gamma}$ is accurately measured~\citep{Planck:2019nip,Planck:2018jri,Aghanim:2018eyx}, the present-day fraction of the energy budget of the Universe in gravitational radiation (\textit{i.e.}, the ratio between the present GW energy density $\rho_{\rm GW}$ and the critical density, $\rho_c=3H^2/8\pi G$), can be easily computed by integrating the spectrum over all scales~\citep{Maggiore:1999vm,Boyle:2005se,Guzzetti:2016mkm}
\begin{equation}
\Omega_{\mathrm{GW}}=\frac{1}{12 H_{0}^{2}} \int \text{d} \ln k \, \mathcal P_{\rm T}(k) \, \dot{T}(\eta_0,k)^2   
\label{Eq:full_OmGW}
\end{equation}
where the contribution of each mode is weighted by (the time derivative of) the so-called transfer function 
\begin{equation}
T(\eta,k)=\frac{h_k(\eta)}{h_k(\eta_i)}
\end{equation}
that takes into account the different time-evolution of modes with different $k$ according to Eq.\eqref{eq:motion}. Assuming that inflation is followed by a standard Hot Big Bang Theory evolution, (\textit{i.e.}, by radiation, matter, and dark energy dominated epochs), the transfer function admits relatively simple semi-analytic solutions and we can estimate the present time contribution at generic frequency $f=k/2\pi$ as~\citep{Akrami:2018odb,Bartolo:2016ami,Cabass:2015jwe,Stewart:2007fu,Graef:2018fzu,Liu:2015psa} 
\begin{equation}
\Omega_{\mathrm{GW}}(f) \simeq\frac{\mathcal P_{\mathrm{T}}(f)}{24 z_{\mathrm{eq}}} 
\label{OmegaGW}
\end{equation}
with $z_{\mathrm{eq}}\simeq 3400$ the redshift at equivalence and $\mathcal P_{\mathrm{T}}$ the spectrum of primordial tensor modes. By using Eq.\eqref{OmegaGW}, putting everything together, we finally get~\citep{Maggiore:1999vm}
\begin{equation}
\Delta N_{\rm eff}^{\rm GW} \simeq \frac{h_0^2}{5.6\times10^{-6}}\left(\frac{1}{24\,z_{\rm eq}}\right) \int_{f_{\rm min}}^{f_{\rm max}} \frac{\mathrm{d}f}{f}\, \mathcal P_{\rm T}(f)
\end{equation}
recovering the standard result that Gravitational Waves contribute to the effective number of relativistic species through the logarithmic integral of their power spectrum over frequencies.

\section{Updated BBN bounds on Inflation}
\label{sec.Appendix-B}

\begin{table*}
	\begin{center}
		\renewcommand{\arraystretch}{1.5}
		\resizebox{0.6\textwidth}{!}{\begin{tabular}{l c c  c  c}
		\hline \hline
		    \textbf{\nq{Parameter \\}} &  \textbf{\nq{BBN-A\\($Y_p + D/H$) }}  &  \textbf{\nq{BBN-B\\($Y_p + \Omega_b\,h^2$)}} &  \textbf{\nq{BBN-C\\ ($Y_p + D/H + \Omega_b\,h^2$)}} \\
		    \hline\hline
			$\Omega_{\rm b} h^2$ &$0.02234\pm 0.00017$&$0.02240\pm 0.00010$&$0.022382\pm 0.000086$\\
			$Y_p$  &$0.24558\pm 0.00010$&$0.24561\pm 0.00010$&$0.245591^{+0.000015}_{-0.000060}$\\
			$(D/H)\cdot 10^{-5}$ &$\,2.527\pm 0.030$&$2.516\pm 0.020$&$2.519\pm 0.016$\\
			$\Delta N_{\rm eff}$ &$< 0.33  \,\, (<0.40)$ &$< 0.32  \,\, (<0.40)$ &$< 0.16  \,\, (<0.21)$\\
			\hline
			\multicolumn{4}{c}{}\\
		    \multicolumn{4}{c}{\textbf{Constraints on Inflation inferred by assuming \autoref{PL}}}\\
		    \multicolumn{4}{c}{}\\
  	        \hline
			$n_{\rm T}$ &$<0.324 \,\, (<0.376)$&$< 0.323 \,\, (<0.374)$&$<0.32 \,\, (0.368)$\\
			$r$ &$<0.037$&$< 0.037$&$< 0.037$\\
			\hline
			\multicolumn{4}{c}{}\\
		    \multicolumn{4}{c}{\textbf{Constraints on Inflation inferred by assuming \autoref{PL2}}}\\
		    \multicolumn{4}{c}{}\\
  	        \hline
  	        $n_{\rm T}$ &$< 1.80\,\,(\text{unc.})$&$ < 1.80\,\,(\text{unc.})$&$< 1.80\,\,(\text{unc.})$\\
			$r$ &$<0.037$&$< 0.037$&$< 0.037$\\
			\hline	\hline
		\end{tabular}}
	\end{center}
	
	\caption{\small Results inferred from BBN primordial abundances. The constraints on $\Omega_{\rm b} h^2$ , $Y_p$ and $10^{5}\cdot(D/H)$ are given at 68\%CL while the upper bounds on $\Delta N_{\rm eff}$ and $n_{\rm T}$ are given at 95\% CL (99\% CL). The horizontal lines divide the constraints on the BBN parameters (that are not sensitive to the model of inflation) from those inferred for the inflationary parameters under the two different parameterizations of the spectrum indicated in the table. A BK18 prior ($r<0.037$ at 95\% CL) is assumed on the tensor amplitude.}
	\label{tab.BBN}
\end{table*}

\begin{figure*}
 	 \centering
 	 \includegraphics[width=0.94\linewidth]{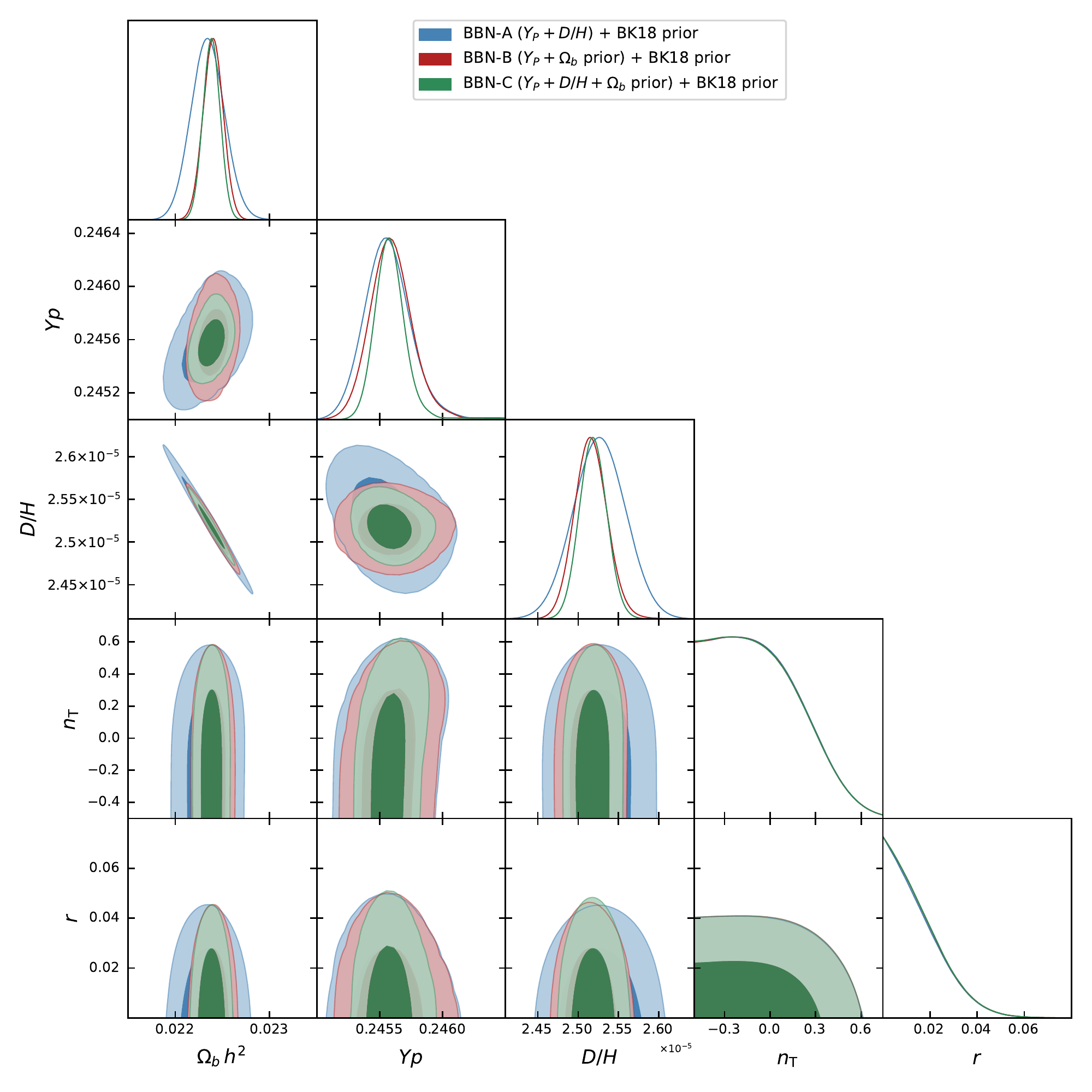}
 	 \caption{\small Two-dimensional 68\% and 95\% CL allowed regions and one-dimensional probability posterior distributions for the most relevant cosmological parameters obtained under the assumption of a power-law spectrum, Eq.\eqref{PL}. The different colors refer to the different data combinations here considered for BBN analyses, see \autoref{tab.BBN}.}
 	 \label{fig:figure6}
\end{figure*}

\begin{figure}
 	 \centering
 	 \includegraphics[width=0.8\columnwidth]{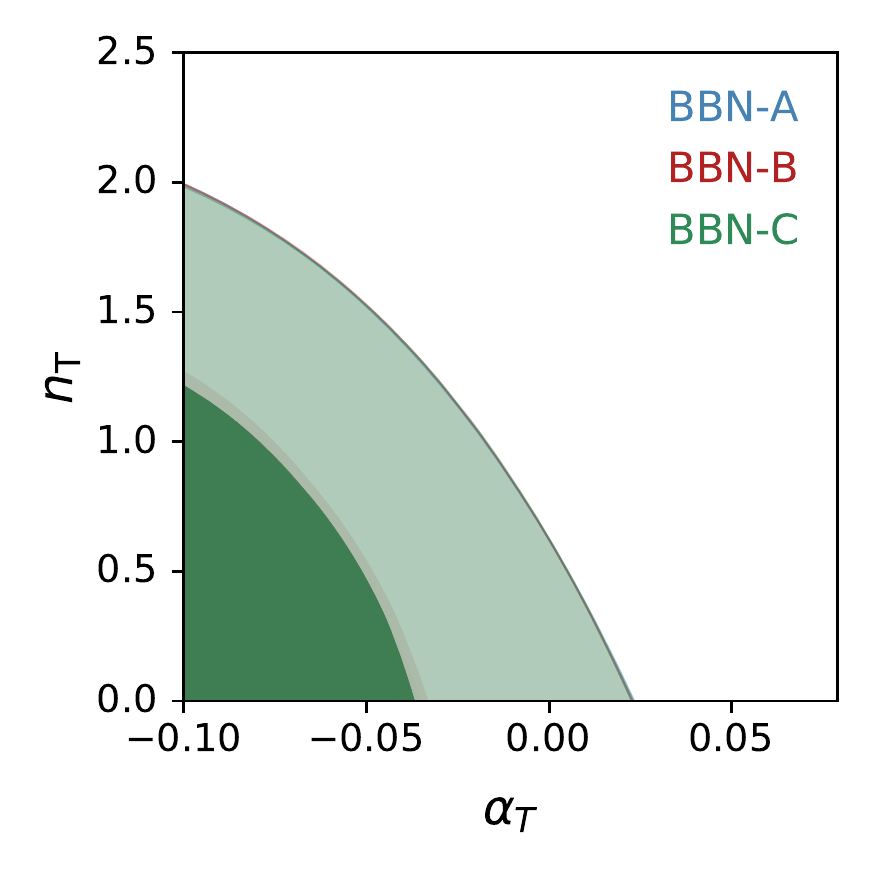}
 	 \caption{\small 2D joint marginalized contours in the $(\alpha_{\rm T}\,,\,n_{\rm T})$ plane obtained by allowing a non-vanishing running $\alpha_{\rm T}=d n_{\rm T} / d\log k$ to vary, see Eq.\eqref{PL2}.}
 	 \label{fig:figure7}
\end{figure}

To enrich and support the analysis carried out in the manuscript, we devote this appendix to the detailed study of the observational constraints on blue-tilted models inflation resulting from the Big Bang Nucleosynthesis epoch. Our aim is twofold: we first update the state-of-the-art results in light of the most recent cosmological observations. Then, retracing the discussion of \autoref{sec.3}, we quantify how such results change with the parameterization of the primordial tensor spectrum.

We start recalling that the Big Bang Nucleosynthesis~\citep{Alpher:1948ve} explains the formation of the first light nuclei heavier than the lightest isotope of hydrogen by a solid understanding of the nuclear interactions involved in their production. It also provides a natural arena to test and constrain extensions to both cosmology and fundamental physics since any proposed model of the early Universe must be able to explain the abundances of light elements inferred by astrophysical and cosmological observations. The reason why the BBN constraining power can be applied to the analysis of blue-tilted models of inflation is quite straightforward: according to the Friedmann equation, additional gravitational radiation (that we parameterized in terms of $\Delta N_{\rm eff}^{\rm GW}$) will increase the expansion rate of the Universe $H(z)$. A faster expansion leads to a higher freeze-out temperature of the weak interactions, implying a higher fraction of primordial Helium and Deuterium, as well as a higher fraction of other primordial elements. This makes BBN an extremely powerful and quite general tool for constraining the total amount of relativistic species in the Universe, with several implications for physics beyond the Standard Model~\citep{Kawasaki:2004qu,Steigman:2007xt,Cyburt:2004yc,Sabti:2019mhn,DEramo:2022nvb}, the Neutrino flavor physics and, in our case, the inflationary cosmology. 

It is instructive to start our analysis by assuming a power-law primordial spectrum given by Eq.~\eqref{PL}.  This simple parameterization has the benefit that all the models are described only by two quantities: the amplitude $r$ and the tilt $n_{\rm T}$. We randomly sample $N=10^6$ linearly distributed values of the amplitude and the tilt in the ranges $r\in[0\,,\,0.1]$ and $n_{\rm T}\in[-2\,,\,2]$, respectively. For each of these points, we compute the contribution to the effective number of relativistic species $\Delta N_{\rm eff}^{\rm GW} (r\,,\,n_{\rm T})$ by Eq.~\eqref{Int1}. Finally, we randomly sample $N$ values of the baryon energy density in the range $\Omega_b h^2\in\left[0.020\,,\,0.025 \right]$ and create a grid in the plane $(\Delta N_{\rm eff}^{\rm GW} ,\Omega_b h^2)$ similar to those usually obtained within the Monte Carlo methods. Then we solve numerically the set of differential equations that regulate the BBN nuclear interactions in the primordial plasma~\citep{PitrouEtal2018,Pisanti:2007hk,Consiglio:2017pot,Gariazzo:2021iiu}. To do so, we made use of the code \textsc{PArthENoPE}~\citep{Gariazzo:2021iiu}. Fixing the values of the neutron lifetime\footnote{The neutron lifetime is fixed to $\tau_n = 879.4$ s, corresponding to the latest measurement reported by the Particle Data Group ($\tau_n = 879.4 \pm 0.6 $ s)~\citep{ParticleDataGroup:2020ssz}}, for each point in the $(\Delta N_{\rm eff} ,\Omega_b h^2)$ plane the code computes the corresponding value of the primordial Helium fraction $Y_P$, the Deuterium abundance $D/H$ and all the other light element abundances. In this way, we can directly compare the results with the values inferred by astrophysical and cosmological observations.  In this regard, our baseline dataset for the BBN analyses consists of:

\begin{itemize} [leftmargin = *]
\item Two independent measurements of the primordial Helium \textit{fraction}, $Y_p = 0.2449 \pm 0.0040$~\citep{Aver:2015iza} and $Y_p = 0.2446\pm 0.0029$~\citep{Peimbert:2016bdg}. 

\item A percent determination of the primordial Deuterium abundance $D/H=\left(2.527\pm0.030\right)\cdot 10^{-5}$ based on six high precision and homogeneously analyzed $D/H$ measurements from~\citep{Cooke:2017cwo}.

\item The value of the baryon energy density parameter $\Omega_b\,h^2=0.0224 \pm 0.0001$ from the final 2018 Planck data release of temperature and polarization CMB angular power spectra~\citep{Aghanim:2018eyx}.

\item A prior on the tensor amplitude $r<0.037$ at 95\% CL coming from a combination of the final 2018 Planck data release of temperature and polarization CMB angular power spectra~\citep{Aghanim:2018eyx} and the B-modes 2018 likelihood from the Bicep Collaboration ~\citep{BICEP:2021xfz}.
\end{itemize}

We apply these priors on the BBN abundances, re-weighting the contributions of the points by means of an "\textit{importance sampling}" statistical method as done in Ref.~\citep{DEramo:2022nvb}. Consequently, we obtain informative posterior distributions for the most interesting parameters to be inferred by observations. We summarize the results in \autoref{tab.BBN} while \autoref{fig:figure6} provides the marginalized posterior distributions of parameters. 

We start by adopting a prior knowledge of the total amount of the primordial Helium $Y_p$ and Deuterium $D/H$ from direct astrophysical measurements, together with a prior on the tensor amplitude from the BK18 likelihood for B-modes polarization. Therefore in this case the free parameters of the sample to be inferred by observations are the baryon energy density and the tensor tilt (the last one to be inferred by the total amount of extra radiation $\Delta N_{\rm eff}$). We refer to this dataset as "BBN-A". From it we derive an upper limit on the additional radiation allowed during BBN epoch of $\Delta N_{\rm eff}<0.3$ at 95\% CL ( $\Delta N_{\rm eff}<0.4$ at 99\% CL), in perfect agreement with the previous results discussed in the literature~\citep{Aver:2015iza,Peimbert:2016bdg,Cooke:2017cwo,DEramo:2022nvb,Giare:2021cqr,Aich:2019obd}. Assuming all this contribution to be made of primordial gravitational waves, we infer an upper limit on the tensor tilt $n_{\rm T}<0.3$ at 95\% CL ( $n_{\rm T}<0.4$ at 99\% CL), which is in line with what we argued in \autoref{sec.3} and with the most recent CMB-analyses \citep{Galloni:2022mok}. 

We test the robustness of our result by considering different combinations of data. In particular, we now impose a prior knowledge on the baryon-energy density $\Omega_b h^2$ as inferred by the Planck collaboration analyzing the last release of the CMB data~\citep{Aghanim:2018eyx} together with the information on the amount of the primordial Helium $Y_p$. We label this case "BBN-B". The free parameters to be determined now are $D/H$ and $n_{\rm T}$. We find that the constraints on the effective number of relativistic degrees of freedom remain basically unchanged with respect to the previous case and so does the limit on the tensor tilt. Notice that, while in this case, we are more dependent on the physics at the recombination epoch, we are relaxing the bound on Deuterium. So we can use the value inferred for this parameter as a consistency check of our analysis, resulting in a great agreement among the different data-combinations.

Finally, for completeness, we combine all these priors together ($Y_p+D/H+\Omega_b\,h^2 +r$). We refer to this dataset as BBN-C. As already pointed out in Ref.~\citep{DEramo:2022nvb}, assuming all this information leads to an improvement in the constraining power on additional radiation with the limit now reading $\Delta N_{\rm eff}<0.16$ at 95\% CL ( $\Delta N_{\rm eff}<0.21$ at 99\% CL). Interestingly, this improvement is not transferred into the bound on the tensor tilt which in fact remains basically unchanged with respect to the previous cases. The reason underlying this lack of improvement can be easily understood by looking at the black dashed line in \autoref{fig:figure1}. This line represents the contribution to $\Delta N_{\rm eff}$ resulting from a blue-tilted power-law spectrum that exponentially grows for positive $n_{\rm T}$. As evident from the figure, when we are close $n_{\rm T}\sim 0.4$ the line in the plane $(n_{\rm T}\,,\,\Delta N_{\rm s})$ becomes almost horizontal. This means that a variation on the y-axis ($\Delta N_{\rm eff}$) does not produce any significant movement in the x-axis direction ($n_{\rm T}$), explaining why we do not get a more tight limits on the tensor tilt.

Aiming to quantify the impact on the results from having assumed a vanishing running $\alpha_{\rm T}=0$, we repeat the same analysis by allowing $\alpha_{\rm T}$ to vary in the range $\alpha_{\rm T}\in[-0.2\,,\,0.2]$. In this case, we parameterize the primordial spectrum through Eq.\eqref{PL2}. We summarize the results inferred for the different datasets in \autoref{tab.BBN}. 

Clearly, both the bounds on the total amount of additional radiation allowed during the BBN epoch and the results on the primordial light element abundance do not change with respect to the previous analysis since they do not depend on the parametrization adopted for the tensor spectrum. Instead, what changes is the limit that we can infer from these limits on the inflationary parameters. In particular, opening to the running completely relaxes the upper limit on positive tensor tilt. This parameter is now constrained to be $n_{\rm T}<1.8$ at 95\% while it is unbounded at 99\% CL. This is due to the strong degeneracy between the tilt and its running, see also their 2D joint marginalized contours shown in \autoref{fig:figure7}. As already discussed in \autoref{sec.3}, a positive running will amplify the power in gravitational waves on small scales, basically miming the effect of a larger scalar tilt. So when the running becomes positive and relatively large, the tensor tilt is only allowed to be either very close to zero or negative, see \autoref{fig:figure7}. Conversely, when the running acquires negative values it strongly reduces the power in gravitational waves and compensates the effect of a larger $n_{\rm T}$. This is why the bounds on the tensor tilt are more relaxed in the region of negative runnings as clearly visible in \autoref{fig:figure7}. These results are in line with what is argued in \autoref{sec.3} and confirm, one more time, that the parameterization adopted for the tensor spectrum is in fact crucial when extrapolating constraints on blue-tilted models of inflation. 

We conclude by pointing out that if also a running of running is allowed to vary in the sample, the tensor tilt is completely unbounded. By extension, this applies to all the other higher-order parameterizations that involve more than two free parameters. 
\label{lastpage}
\end{document}